\PassOptionsToPackage{sort&compress}{natbib} % reference ranges
\documentclass[preprint,review,10pt,5p,times,twocolumn]{elsarticle}

\usepackage{amssymb}
\usepackage{amsmath}
\usepackage{enumitem} % in the preamble
\usepackage{array}

\usepackage{url}
\usepackage{algorithmic} % added
\usepackage{algorithm}
\usepackage{verbatim} % added
\usepackage{multicol}
\usepackage{multirow}
\usepackage{adjustbox}
\usepackage{setspace}
\usepackage{xcolor}

\newcommand{\edit}[1]{{\color{black}#1}}
\newcommand{\editt}[1]{{\color{black}#1}}

\usepackage{lineno}

\journal{Energy \& Buildings}

\begin{document}

\begin{frontmatter}



\title{Adapting to Change: A Comparison of Continual and Transfer Learning for Modeling Building Thermal Dynamics under Concept Drifts}
%old title: The Benefit of Collecting Data over Time for Building Dynamics Modeling -- A Comparison of Continual Learning and Transfer Learning for a Changing Built Environment


\author[label3,label1]{Fabian Raisch \corref{cor1}}
\ead{fabian.raisch@tum.de}
\cortext[cor1]{Corresponding author}

\author[label2]{Max Langtry}
\author[label3]{Felix Koch}
\author[label2]{Ruchi Choudhary}

\author[label1]{Christoph Goebel \corref{cor2}}
\author[label3]{Benjamin Tischler \corref{cor2}}

\cortext[cor2]{Authors jointly supervised.}

\affiliation[label3]{organization={Technical University of Applied Sciences Rosenheim}, city={Hochschulstraße 1, Rosenheim, 83024}, country={Germany}}
\affiliation[label1]{organization={Technical University of Munich}, city={ Arcisstraße 21, Munich, 80333}, country={Germany}} 
\affiliation[label2]{organization={University of Cambridge}, city={Trumpington Street, Cambridge, CB2 1PZ}, country={UK}}

%% Abstract
\begin{abstract}

%However, it remains unclear how new operational data should be incorporated into the initial, fine-tuned model as this data becomes available over time.
%There has yet to be a comprehensive study of how to incorporate new monitoring data over time to improve prediction accuracy and address the challenges of concept drifts (changes in dynamics) in the built environment.
%regarding different CL and TL strategies for changing built environments is so far missing.

Transfer Learning (TL) is currently the most effective approach for modeling building thermal dynamics when only limited data are available. TL uses a pretrained model that is fine-tuned to a specific target building. However, it remains unclear how to proceed after initial fine-tuning, as more operational measurement data are collected over time. This challenge becomes even more complex when the dynamics of the building change, for example, after a retrofit or a change in occupancy. In Machine Learning literature, Continual Learning (CL) methods are used to update models of changing systems. TL approaches can also address this challenge by reusing the pretrained model at each update step and fine-tuning it with new measurement data.
A comprehensive study on how to incorporate new measurement data over time to improve prediction accuracy and address the challenges of concept drifts (changes in dynamics) for building thermal dynamics is still missing.
Therefore, this study compares several CL and TL strategies, as well as a model trained from scratch, for thermal dynamics modeling during building operation. The methods are evaluated using 5--7 years of simulated data representative of single-family houses in Central Europe, including scenarios with concept drifts from retrofits and changes in occupancy. We propose Seasonal Memory Learning \edit{(SML), a CL strategy} that provides greater accuracy improvements than existing CL and TL methods, while maintaining low computational effort. SML outperformed the benchmark of initial fine-tuning by \edit{42.5}\% without concept drifts and \edit{48.3}\% with concept drifts.

\end{abstract}

%

%% Keywords
\begin{keyword}
Continual Learning \sep Transfer Learning \sep Building Thermal Dynamics \sep Data-Driven Model \sep Data Requirements  \sep Concept Drift\edit{\sep Retrofit \sep Change in Occupancy}
\end{keyword}

\end{frontmatter}

\section{Introduction}

%% <Max> Currently this is very focused on methods; my preference is to describe the underlying problems that need to be tackled, and then discuss how existing methods address them, and their limitations.
Building energy usage accounts for approximately one-third of global greenhouse gas emissions \cite{eea2023DecarbonisingHeatingCooling}. Advanced control systems, such as Model Predictive Control (MPC), can reduce these emissions by 10–50\% \cite{Drgona.2020, Hilliard02072016}. MPC requires an accurate model of the underlying building dynamics. Controllers based on Reinforcement Learning (RL) schemes can also benefit from a high-accuracy dynamics model \cite{CORACI2023117303, YU2022109458, ZOU2020106535}. Another option to increase energy efficiency in buildings is to employ fault detection and diagnosis systems (FDD). FDD systems use a dynamics prediction model to compare operational data to detect faults \cite{DU20091624, CHEN2023121030}. Hence, developing accurate thermal dynamics models is crucial for improving building energy efficiency. For both control and FDD applications, data-driven methods are a promising solution for learning these models \cite{Drgona.2020, CHEN2023121030}. Their main advantage lies in avoiding manual modeling, thus enabling large-scale deployment of energy-efficient systems. In this context, Machine Learning (ML) techniques have received significant attention. These methods have the advantage of learning complex, nonlinear system behavior directly from data without requiring explicit knowledge of the particular target building. However, one major drawback of ML models in buildings is the need to have sufficient, representative training data to achieve good prediction accuracy. Currently, multiple months to years of data are required for good model performance \cite{choi_performance_2023, pinto_sharing_2022}. However, in many building energy systems, little monitoring data is available at the time control or FDD systems are installed. Transfer Learning (TL) can overcome this issue of limited data availability, as it uses a pretrained model as a starting point for modeling the target building. This reduces the data requirements to a few months or weeks to achieve comparable performance \cite{pinto2022transfer, peirelinck2022transfer}. 
%Due to its recent success, TL can be considered the state-of-the-art for modeling building thermal dynamics for the initial phase of building operation. 
%This is especially true since the authors of \cite{raisch2025gentlgeneraltransferlearning} solved the issue of the source selection problem by offering a universal TL model for building thermal dynamics. This approach constantly achieved better prediction performance than a model from scratch. 

% key research question
TL is currently the best method for creating building thermal dynamics models at the initial phase of modeling \cite{raisch2025gentlgeneraltransferlearning, pinto_sharing_2022, li_building_2024}. However, as buildings operate, data on the particular behavior of the target building is collected over time. This data can be used to iteratively update the model, a process we refer to as \textbf{adaptive learning}. Furthermore, the building may enter operating states that are not well represented in the initial fine-tuning data. This issue is particularly pronounced when less than one year of data is available, as the data does not cover all seasons, and train-validation splitting further reduces usable training data \cite{LANGTRY2024114605, pinto_sharing_2022}. As a result, model accuracy may degrade over time, which can reduce its performance in the considered application \cite{choi_performance_2023, grubinger_generalized_2017, Drgona.2020}. This leads to a key research question, 
%``Should building dynamics models be updated after initial fine-tuning as more data becomes available, and if so, what is an appropriate updating interval?''. impact of incorporating additional
%``What is the impact of incorporating new measurement data  in building dynamics models on the prediction accuracy, and how does the update frequency used for updating affect model performance?''
``How does incorporating new measurement data into building dynamics models affect prediction accuracy after initial fine-tuning, and how does the update frequency influence model performance?''
%This may be particularly relevant as building dynamics change over time, for example, due to seasonal variations. %
%To address this, data-driven models should be continuously adapted to reflect the current behavior of the building. 
%This leads to a second question, ``How can newly collected data be incorporated into the model in a systematic and effective manner?''.
%Should building dynamics models be updated after initial fine-tuning as more data becomes available over time?
%However, the question persists of how to proceed from there during building operations as more and more data becomes available over time. This is especially challenging since buildings may change over time. Especially due to seasonal changes, the building enters states that have not necessarily been modeled well. This effect may especially occur in the case when less than one year of data has been collected. In the end, this can lead to worse control performance in the building \cite{choi_performance_2023, grubinger_generalized_2017}. For this reason, data-driven models should be adjusted during building operations to current conditions. This leads to the question of how to encompass the newly collected data during building operations. 
\begin{figure}[b]
    \centering
    \includegraphics[width=1 \linewidth, trim={0.7cm 0 0.7cm 0},clip]{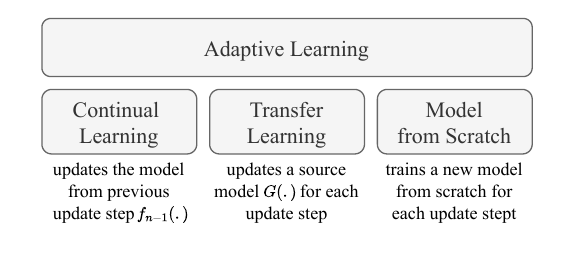}
    \vspace*{-1.2cm}
    \caption{\edit{Overview of adaptive learning approaches.}}
    \label{fig:adapt_learning_overview}
\end{figure}

% Methods we compare
Several strategies exist for updating data-driven models during operation. \edit{Figure \ref{fig:adapt_learning_overview} summarizes the different adaptive learning approaches.} A common approach is to train a model from scratch once sufficient data has been collected \cite{pinto_sharing_2022, choi_performance_2023}. 
%For example, \cite{pinto_sharing_2022} compared a TL-based model with a model trained from scratch in an adaptive learning setting for thermal dynamics in buildings. They recommend a model from scratch after one year of data is available. However, their study is highly limited to only one specific source to target pair, and one update strategy.
An alternative approach is to use Continual Learning (CL) techniques from the ML literature \cite{verwimp_continual_nodate, CL_review}. CL continuously updates the model over time as new data is collected.
%\edit{Adaptive Learning, Continual Learning, and Transfer Learning are closely related concepts that are often used interchangeably in the literature. Table \ref{...} summaries the relationship between these terms and how they will be used in this paper.}
There are several different \edit{CL} mechanisms for doing this update. A key challenge in CL is balancing catastrophic forgetting (the loss of previously learned knowledge when adapting to new data \cite{noauthor_overcoming_nodate}) and plasticity (the model's ability to integrate new information \cite{berariu2021study}). Research on CL in the context of building thermal dynamics modeling remains scarce (as discussed in Section \ref{ch:literature_review}), making the assessment of suitable methods an open challenge.
%Hardly any studies have investigated CL in the context of building thermal dynamics modeling (cf. Section \ref{ch:literature_review}). 
%However, these studies are very limited to specific use cases (cf. Section \ref{ch:literature_review}).
A third alternative is to adapt TL methods for adaptive learning. For instance, \cite{raisch2025gentlgeneraltransferlearning} proposes a general, pretrained TL model. This model can be fine-tuned specifically for the current operational conditions by using exclusively newly measured data. Alternatively, one can apply all accumulated data over time to fine-tune the general TL model in each update step. This approach is similar to training a model from scratch, with the difference that the general pretrained model is used for weight initialization.
All of these strategies can lead to differences in model performance over time and computational update cost. Computational cost is a critical consideration in building operation, as on-site hardware often has limited processing capabilities \cite{Drgona.2020}.
%There are several different strategies, like training a model from scratch as enough data becomes available. This is suggested from \cite{pinto_sharing_2022} after one year of data becomes available. They compared a TL model versus a model from scratch, but only considered one building and one updating strategy for this study. An alternative would be to use strategies from Continual Learning (CL) \cite{verwimp_continual_nodate, lopez-paz_gradient_2022}. CL consistently updates the data-driven model, but can employ different strategies to do so. In this domain, the trade-off between catastrophic forgetting and plasticity is of interest. Catastrophic forgetting describes the effect of overfitting to newly collected data but forgetting patterns of old data \cite{noauthor_overcoming_nodate}. Plasticity, on the other hand, describes the ability of an ML model to learn on new data \cite{berariu2021study}. Few studies have investigated CL for building thermal dynamics \cite{ZAMORAMARTINEZ2014162, pinto_sharing_2022}. However, these studies are very limited to specific use cases. A third alternative would be to reinvent TL strategies. As the authors of \cite{raisch2025gentlgeneraltransferlearning} proposed a general pretrained TL model, this can be employed severely in a adaptive learning setting: The newly incoming data can be used to fine-tune the general model for this specific season. Alternatively, all available data can be utilized to fine-tune the general TL model in an accumulative way. 
%RQ1:
This raises the question, ``Which update strategy is most suitable as more data become available, particularly regarding the trade-off between model performance and computational costs in building thermal dynamics?''
%The question arises: which of the state-of-the-art CL algorithms performs best in the target building with respect to a pretrained model? Additionally, it is unclear whether it makes sense to reinitialize the target's model with a TL model when it comes to sudden concept drifts. 
%Building dynamics can change due to factors beyond seasonal variations. This includes concept drifts caused by occupant changes or retrofits \cite{li_large-scale_2023, deng_2022_drift}. These events create a lasting impact on the dynamics, unlike weather patterns, which evolve gradually and may become irrelevant over time. Li et al. \cite{li_large-scale_2023} analyzed concept drifts on electrical load forecasting in buildings. They found Elastic Weight Consolidation and Gradient Episodic Memory to work best for their application. However, they could not attribute the observed drifts to specific events. Additionally, electrical load forecasting differs fundamentally from thermal dynamics prediction due to longer time lags and thermal inertia. The authors therefore emphasized the need to study concept drifts for building thermal modeling. 

Building dynamics can change over time due to several factors. One reason is seasonal weather variation (\textbf{feature drifts}), as a model might be fitted to winter data but is applied to summer data \cite{dirfts_gama}. Other reasons are retrofits or changes in occupancy (\textbf{concept drifts}) \cite{dirfts_gama, deng_2022_drift}. Concept drifts create a lasting impact on the dynamics (conditional distribution of the output given the input), unlike feature drifts, which affect only the statistical distribution of the input features without altering the underlying system behavior. These drifts can have a major impact on model performance, highlighting the necessity of investigating concept drifts in building thermal dynamics modeling, as emphasized in \cite{li_large-scale_2023}. Similarly, Choi et al. \cite{choi_performance_2023} underlined the need for adaptive learning to account for changing building dynamics, such as retrofits.
%RQ3:
This raises the research question, ``How do drifts in the built environment influence the modeling of thermal dynamics, and which strategies best adapt models to these evolving conditions?''

\subsection{Related Work}
\label{ch:literature_review}

Adaptive learning for building thermal dynamics models has received significant attention in recent years. Many studies employ gray-box resistance-capacitance network models. Alongside classical gray-box approaches, such as \cite{7551202}, ML-enhanced gray-box methods have emerged \cite{xie_lifelong_2024}.
For example, Dong et al. \cite{10.1145/3307772.3331029} propose a hybrid adaptive thermal dynamics prediction model that combines a parametric component with a nonparametric random forest model for commercial buildings. Their approach achieves more than twice the accuracy of a baseline persistent model. They suggest extending the method to residential buildings and exploring alternative regression algorithms in future work.

In contrast, many studies focus on pure black-box solutions that require no prior knowledge of the system \cite{raisch2025gentlgeneraltransferlearning, li_building_2024, choi_performance_2023, Martinez2020LSTMMLP, Mtibaa2020LSTM, ZAMORAMARTINEZ2014162}. Zamora-Martínez et al. \cite{ZAMORAMARTINEZ2014162} demonstrate the feasibility of adaptive learning for indoor temperature prediction. They compared an artificial neural network with a Bayesian model in an online setting using different hyperparameters and found that simpler models often outperformed more complex ones. However, the generalizability of the study is limited, as they utilize only one building and two small datasets of 28 and 14 days, as well as only one updating method.
Adaptive learning can also be embedded within the temperature control policy using Reinforcement Learning (RL). Coraci et al. \cite{CORACI2023120598} compare adaptive Transfer Learning (TL)-RL, which uses a pretrained policy from a source building, with adaptive RL, which learns from scratch without prior knowledge of the source building. Tests on 19 target buildings show that TL-RL enables faster training and better performance.  
None of the above studies address computational demands of updates or consider concept drifts in the building. In addition, the studies use only a single update strategy. 

The rapidly expanding field of Continual Learning (CL) offers a wide range of advanced adaptive learning approaches. 
For example, Lopez-Paz \& Ranzato \cite{lopez-paz_gradient_2022} present an algorithm that employs episodic memory to balance the loss between current and past updates. Kirkpatrick et al. \cite{noauthor_overcoming_nodate} identify network parameters that are important for previous updates and penalize changes to these parameters during updates to reduce catastrophic forgetting. 
In building thermal modeling, modern CL strategies are poorly addressed so far. One reason may be that most available datasets covering the thermal side include only a few buildings and short time spans, often no more than one year \cite{present2024resstock, nweye2025citylearn2, pullinger2021ideal, ecobee}. \edit{Additionally, real-world datasets are quite rare that include all desired feature variables, like indoor temperatures and control signals.} In contrast, datasets for building energy or load measurements cover larger numbers of buildings and longer periods \cite{miller2020buildinggenome2, emami2023buildingsbench}.
For load prediction, Zhou et al. \cite{zhou_elastic_2022} propose a CL approach based on Elastic Weight Consolidation (EWC). They assume to have one year of historical data available to train a base model. The base model is updated with EWC incrementally over time. The EWC approach outperforms the standard CL techniques, Accumulative Retraining (AR) and Incremental Learning (IL), and requires considerably less training time. However, their study relies on one year of data for pretraining the base model, which is generally unavailable for most buildings. The authors acknowledge that reducing the amount of pretraining data could limit the method’s performance. Additionally, the study focuses on only one building, limiting the scope of the study to other buildings.
Within building load forecasting, Li et al. \cite{li_large-scale_2023} present a more comprehensive study. Apart from EWC, AR, and IL, they investigated more recent regularization-based and memory replay-based methods. They also analyze the methods’ ability to handle concept drifts in time series data. Their results show that Gradient Episodic Memory (GEM) and EWC perform best overall, while IL (updating all network parameters only on new data) handles sudden concept drifts best. The study focuses exclusively on CL strategies for only 2 years of data and does not include models trained from scratch or TL approaches. Additionally, the causes of the observed drifts remain unclear, as they could not link the drifts to specific events like retrofits or equipment performance degradation.
In addition, the applicability of these findings to building thermal dynamics remains unclear. Thermal dynamics modeling is more complex due to long time lags, thermal inertia, and the need for higher temporal resolution \cite{Drgona.2020}, unlike the one-hour cumulative electrical load prediction as in \cite{li_large-scale_2023}. Consequently, Li et al. call for further research focused on thermal modeling. 

% TL 
Apart from the RL study \cite{CORACI2023120598}, none of the previous studies considered a TL model in the adaptive learning framework.
Pinto et al. \cite{pinto_sharing_2022} compared adaptive learning from scratch with adaptive TL. Their results show that the performances of TL and learning from scratch converge over time, with the latter outperforming TL after a year of new data. They recommend starting with adaptive TL and transitioning to adaptive learning from scratch once sufficient target data accumulates. However, their recommendation does not consider limited computational resources as well as more advanced updating methods. They also note that a TL model that generalizes better might lead to different results. Furthermore, the study is limited to a single source-target pair and does not address concept drifts.

\subsection{Research gap and contribution}
\label{ch:gep}

%Our literature review reveals a research gap in the domain of adaptive learning applied to thermal dynamics modeling for buildings. Most studies concentrated on building load prediction rather then thermal dynamics modeling. No studies have investigated concept drifts within the building and their effect on thermal dynamics modeling. Additionally, a comprehensive study that compares CL and TL strategies as well as a model from scratch in an adaptive learning setting is so far missing. Lastly, all studies focus solely on prediction performance and do not account for computational resources.
Our literature review reveals a research gap in the domain of adaptive learning applied to thermal dynamics modeling for buildings. \edit{According to \cite{li_large-scale_2023, choi_performance_2023}, the effects of concept drifts arising from seasonal changes, retrofits, or changes in occupancy on the accuracy of adaptive thermal models have not yet been examined.
Additionally, most studies focus exclusively on CL methods (as in \cite{li_large-scale_2023, zhou_elastic_2022, ZAMORAMARTINEZ2014162}) or TL methods (as in \cite{pinto_sharing_2022, CORACI2023120598}) but do not provide a comprehensive comparison of several adaptive learning strategies.
Finally, most existing studies focus on prediction accuracy but neglect the computational trade-offs that determine the practical feasibility of deploying adaptive models on resource-constrained devices.
}

To address this research gap, we introduce a comprehensive study of adaptive learning for thermal dynamics modeling. We evaluate several CL algorithms, TL strategies, and a model from scratch to investigate how to best use newly collected data during building operation. Building on these insights, we propose two new, tailored variants of the CL and TL methods. These are Seasonal Memory Learning (SML) and event-based Accumulative Learning on General Model (eALG). \edit{Especially, SML outperformed the remaining approaches for the drifted and undrifted scenarios, closely followed by eALG.} We further compare all methods in terms of prediction accuracy across different update intervals and their computational requirements.
We investigate three scenarios: buildings that are solely affected by seasonal variations, buildings that are affected by a retrofit, and buildings that are affected by a change in occupancy. For demonstration, we use 8 target buildings over a period of five years. Additionally, we perform a fourth scenario, the large-scale analysis, that investigates 40 different buildings with individual drifting schedules over a period of seven years. The selected buildings display varying building properties, locations, and occupancy profiles representing the distribution of single-family houses within Central Europe. With the results from this work, we can provide recommendations for managing long-term thermal dynamics models, which can directly contribute to more energy-efficient operation in buildings. 

The key contributions of this study are:
\begin{itemize}
    \item We examine how to most effectively use incoming monitoring data to enhance building thermal dynamics models during operation. % We show how to encompass new data for thermal dynamics modeling as it is acquired over time
    \item This is the first study to comprehensively investigate the impact of seasonal variations, retrofits, and changes in occupancy on building thermal dynamics models.
    \item We provide the first systematic comparison of Transfer Learning (TL) models, Continual Learning (CL) models, and models trained from scratch in an adaptive learning context, evaluating both prediction performance and computational efficiency.
    %\item We propose two novel adaptive learning approaches -- one based on CL and one based on TL -- which achieve the best predictive performance in our evaluation.
    \item \edit{We propose Seasonal Memory Learning, a CL approach that achieves the best prediction performance in our evaluation for drifted and undrifted scenarios.}
\end{itemize}

The remainder of the paper is structured as follows. Section \ref{ch:method} details the research methodology of the paper. Section \ref{ch:results} presents the experiments and Section \ref{ch:discuss} discusses the results. Finally, we draw a conclusion in Section \ref{ch:conclusion}.

\section{Methodology}
\label{ch:method}

\begin{figure*}[h!]
    \centering
    \includegraphics[width=\linewidth]{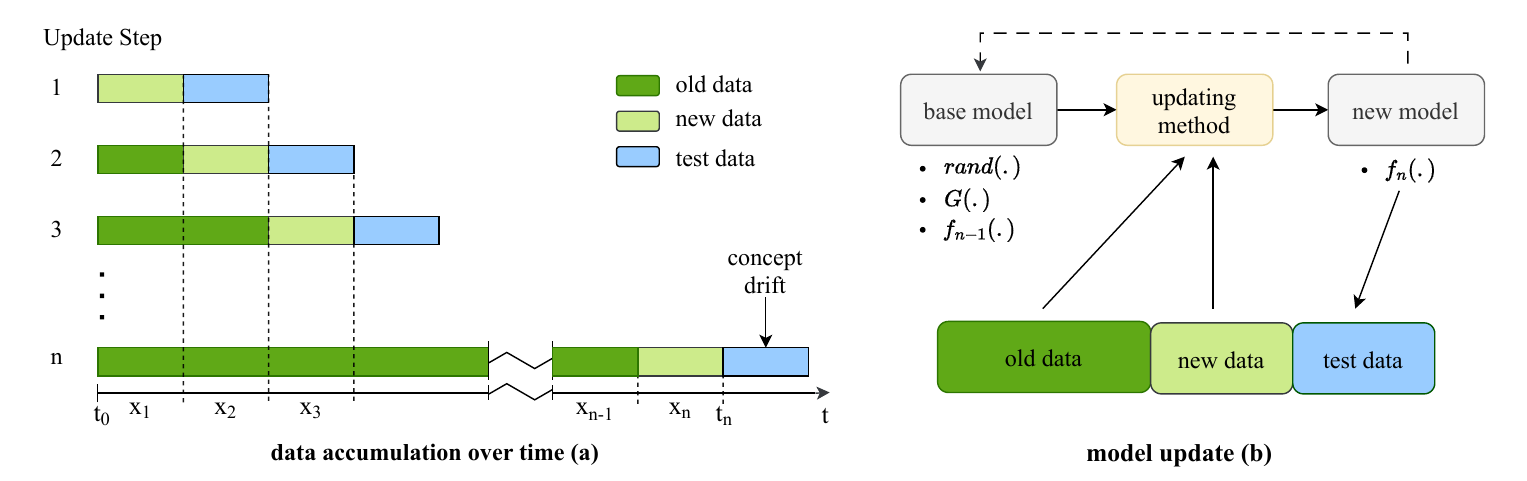}
    %\vspace*{-0.75cm}
    \caption{Illustration of the adaptive learning setup. Base models are a randomly initialized model to train a model from scratch ($rand(.)$), the general TL model ($G(.)$), or the model from the previous update step ($f_{n-1}(.)$).}
    \label{fig:setup}
\end{figure*}

This study investigates thermal dynamics modeling during building operation. We follow the approach according to Figure \ref{fig:setup}. Figure \ref{fig:setup} (a) illustrates the data accumulation over time for
one target building. Modeling of one target building starts at some time instance $t_0$. Thereafter, the building evolves and continuously collects new data that can be used for updates. Figure \ref{fig:setup} (b) shows the basic procedure for the model update, which we refer to as adaptive learning.

In the following, we describe the data simulation and the specifications for the considered buildings in Section \ref{ch:data}. Section \ref{ch:concept_drifts} explains concept drifts and how we employed them in the data generation. In Section \ref{ch:large-scale}, we present an additional building dataset that we use for a large-scale analysis. Next, we cover the general concepts of thermal dynamics modeling in Section \ref{ch:pretrain}. In Section \ref{ch:CL}, we explain the adaptive learning framework and how to apply it to the target buildings. Finally, Section \ref{ch:evaluation} describes the evaluation of the adaptive learning methods.

%In this study, we investigate thermal dynamics modeling during building operation, where several types of concept drift may occur. The choice of data is decisive for the analysis, as it determines both the nature of the drifts and the evaluation of the proposed methods. For this reason, we first provide a detailed explanation of the building data and its drifts in the Sections \ref{ch:data},
%Therefore, we consider the approach according to Figure \ref{fig:setup}. Figure \ref{fig:setup} (a) shows the stream of incoming data over time. The modeling of one specific target building starts at some time instance $t_0$. Thereafter, the building evolves and continuously collects new data, which can be used for updates. After the first update, data from previous intervals is available. Depending on the chosen updating strategy, both new and old data may be used for the model update to obtain the new model, as illustrated in Figure \ref{fig:setup} (b). The new model is then applied to the next consecutive time interval for the intended application (FDD or control). During operations, we may observe concept drifts according to several factors. The following section will explain the buildings' data considered in the study.

\subsection{Building simulation}
\label{ch:data}

To robustly investigate adaptive learning during building operation, we require data from multiple buildings. The data also need to include concept drifts, as these are part of the investigation. In our case, these will be retrofits and changes in occupancy, as further explained in Section \ref{ch:concept_drifts}. Additionally, the data should cover a long period of time to determine the long-term success of adaptive learning. \edit{Last, a minimum of 490 building time series is required, as we employ the TL approach according to \cite{raisch2025gentlgeneraltransferlearning} that uses 450 buildings for pretraining and the remaining ones for testing.}
As no real-world dataset provides a sufficiently long observation period with concept drifts across multiple buildings (cf. Section \ref{ch:literature_review}), we simulate building operational data for this study.
We employ the BuilDa simulation framework of \cite{builDaReview}. BuilDa provides the option to generate high-fidelity data of multiple single-family buildings with varying properties. Building parameters can be configured directly in Python, allowing simulations to run without manually altering the underlying simulation model.
The simulation model is implemented in Modelica and exported as a Functional Mock-up Unit (FMU) \cite{blochwitz2012functional}. 
The Modelica model includes a single-zone, two-floor building model with detailed envelope components, windows in all orientations, and an ideal heat source allowing the representation of various heating systems. The heat source nominal power is sized for each simulation based on transmission and ventilation losses \cite{normHeizlast}. The heat source operates with a maximum of the nominal power and employs a proportional controller for indoor temperature control.
The simulation framework was validated according to ANSI/ASHRAE 140-2004 standard using the test cases TC600, TC900, TC600FF, and TC900FF \cite{ASHRAE140}. 

The framework includes an occupancy profile generator, which produces user-specific temperature setpoints as well as schedules for internal heat gains, and window openings with natural ventilation in each building simulation. The generator selects a random number of \editt{residents with a Gaussian distribution $\mathcal{N}(\mu = 2, \sigma^2 = 1)$ and a minimal value of 1. The number of residents influences the schedules for internal gains and window opening based on \cite{ISO_Norm, ainsworth1993compendium}. For the internal gains, the number of residents is multiplied by a noisy standard occupancy profile (provides hourly number of present people for each type of day - workday, holiday, Saturday, Sunday) based on \cite{Pfluegl2022LoadProfileGenerator} with $\mathcal{N}(\mu = \text{standard profile value}, \sigma^2 = 2$). The simulation framework determines the window opening based on a randomly selected occupants’ conscientiousness regarding air quality $\mathcal{N}(\mu = 1, \sigma^2 = 0.3)$ as well as the occupancy, occupant weight, zone area, and metabolic equivalent rate to mimic the zone CO$_2$ concentration.}
%These schedules account for occupant weight, type of day (workday, holiday, Saturday, Sunday), number of occupants, occupants’ conscientiousness regarding air quality, zone area, and metabolic equivalent rate.
The indoor temperature setpoint \editt{is drawn from a discrete uniform distribution between} 20 \textdegree C and 24 \textdegree C \editt{with a step size of 0.5 \textdegree C}. For 70\% of the profiles, an additional night setback was chosen \editt{uniformly at random} between 0.5 \textdegree C and 4 \textdegree C. The remainder of the profiles use no night setback, which results in a constant setpoint. We refer to \cite{builDaReview} \editt{and their GitHub Repository \cite{BuilDaRepo}} for further information regarding the \editt{occupancy profile generation}. Figure \ref{fig:energy_consumption} illustrates the impact of two occupancy profiles (base case \& change in occupancy) on the annual energy consumption for the buildings from Table \ref{tab:8targets}. 

%The framework includes an occupancy profile generator. This results in user-specific temperature setpoints, a schedule for the internal heat gains due to occupancy, and window openings due to natural ventilation for each building simulation. The profile generator selects a random number of occupants between 1 and 5. It generates the schedules for the internal gains and the window opening based on \cite{ISO_Norm, ainsworth1993compendium}, which is based on the weight per person, the type of day (workday, holiday, Saturday, Sunday), the occupancy, the occupants' conscientiousness rate for air quality, the zone area, and the metabolic equivalent rate.
\begin{table}[!b]
    \centering
    \begin{tabular}{ccc} \hline
         \textbf{Parameter}&  \textbf{Unit} & {\textbf{Targets}}  \\ \hline 
         $U$-$value_{wall}$ & $[W/m^2K]$ &  $\{0.25, 0.55, 0.85, 1.15\}$ \\ 
         $c_{wall}$& $[kJ/m^2K]$ & $\{40, 150, 280\}$ \\ 
         $f_{win}$& - & $\{0.16, 0.19\}$  \\ 
         $A_{ground}$& $[m^2]$ & $\{70, 100\}$ \\ 
         Weather & - & Munich, Amsterdam, Bratislava \\ \hline
    \end{tabular}
    \caption{Parameter distribution for the target buildings, adapted from \cite{raisch2025gentlgeneraltransferlearning}, including the insulation level of exterior wall ($U$-$value_{wall}$), area-specific heat capacity of exterior wall ($c_{wall}$), the window size to wall area ratio ($f_{win}$), and the building ground area ($A_{ground}$). The U-values for the windows and the roof are simplified to $U$-$value_{win}=1+U$-$value_{wall}$ and $U$-$value_{roof}=U$-$value_{wall}$, respectively. Remaining building parameters correspond to respective energy efficiency levels depending on the selected values from this table, according to \cite{tabula, ASHRAE140, bottom_ibpsa_2017}.}
    \label{tab:targets_data}
\end{table}

For the parameter configuration of the buildings simulation, we follow the authors of \cite{raisch2025gentlgeneraltransferlearning}.
They used the BuilDa framework to generate data for 450 different source buildings to train a general, pretrained TL model. After pretraining, they tested the general TL model on 8 unseen buildings, namely the target buildings. We reuse their general TL model for thermal dynamics modeling, as explained in Section \ref{ch:pretrain}. Hence, we can only consider target data for testing.
They selected the parameters for the target buildings according to the distribution shown in Table \ref{tab:targets_data}. Table \ref{tab:targets_data} shows the parameters with the greatest influence on the thermal dynamics of the building \cite{Thomas2006Environmental}, with the values selected according to \cite{tabula}. The weather locations are based on cities in Central Europe with a cold-temperate climate, where heating systems are predominantly used, but no cooling systems \cite{schnieders2020design}. The target buildings represent single-family houses built between 1949 and today, covering a substantial share of residential buildings in Central Europe \cite{tabula, TASK442013}. For their study, they used a representative subset from Table \ref{tab:targets_data}, resulting in 8 target buildings, as shown in Table \ref{tab:8targets}. We use the same data for consistency and because it represents a broad distribution of buildings.
For a robust evaluation of adaptive learning, we will additionally perform a large-scale analysis using five times more buildings, as further explained in Section \ref{ch:large-scale}. Therefore, we use a randomly drawn subset of 40 buildings from Table \ref{tab:targets_data}. We select this number to balance representativeness and computational burdens. 

\edit{We also evaluate the proposed methodology by adding artificial noise to the simulated data.} \editt{Noise was added uniformly to all variables according to \cite{choi_performance_2023}.}

\newcommand{\rot}[1]{\multicolumn{1}{c}{\hspace{-1ex}\adjustbox{angle=30,lap=\width-1em}{#1}}}
% alternatively use \rotatebox{90}{<content>}
\begin{table}[!b]
    \centering
    \begin{tabular}{c@{\hskip .33cm}|ccccccc} 
        \rot{Target building} & \rot{$U_{wall}$ $[W/(m^2K)]$} & \rot{\textbf{$c_{wall}$ $[kJ/(m^2K)]$}} & \rot{\textbf{$f_{win}$}}  & \rot{\textbf{$A_{ground}$ $[m^2]$}} & \rot{$T_{sp,\,day}$ [\textdegree C] } & \rot{$\Delta T_{night}$ [\textdegree C]} & \rot{Weather} \\  \hline
        T1 & 0.25 & 40  & 0.16 & 70 & 22.0 & 1.0 & Bratislava \\
        T2 & 0.25 & 280 & 0.19 & 100 & 21.0 & 0.0 & Amsterdam \\  
        T3 & 0.55 & 150 & 0.16 & 70& 23.0 & 0.0 & Amsterdam \\
        T4 & 0.55 & 280 & 0.19 & 100 & 20.5 & 1.5 & Munich \\ 
        T5 & 0.85 & 40 & 0.16 & 70 & 22.0 & 2.5 & Munich \\ 
        T6 & 0.85 & 150 & 0.19 & 100 & 22.5 & 0.5 & Bratislava \\ 
        T7 & 1.15 & 280 & 0.16 & 70 & 23.0 & 0.0  & Bratislava \\ 
        T8 & 1.15 & 40 & 0.19 & 100 & 23.0 & 1.5 & Amsterdam \\ \hline
     \end{tabular}
     \caption{Properties of target buildings, selected to cover distribution from Table \ref{tab:targets_data}, according to \cite{raisch2025gentlgeneraltransferlearning}. $T_{sp,\,day}$ is the daytime temperature setpoint and $\Delta T_{night}$ a potential night setback.}
     \label{tab:8targets}
\end{table}

\subsection{Concept drifts during building operation}
\label{ch:concept_drifts}

%Building dynamics can change over time due to several factors. One reason is seasonal change (feature/virtual drifts), as a model might be fitted to winter data but is applied to summer data \cite{dirfts_gama}. Other reasons are changes caused by occupant changes or retrofits (concept drifts) \cite{dirfts_gama, deng_2022_drift}. Concept drifts create a lasting impact on the dynamics (conditional distribution of the output given the input), unlike feature drifts, which affect only the statistical distribution of the input features without altering the underlying system behavior. These drifts can have a major impact on model performance. 
Building thermal dynamics change for several reasons. The most obvious reason is seasonal variation due to weather, which we classify as \textbf{feature drifts}. Feature drifts define changes in the statistical distribution of the input features without altering the underlying physical properties of the environment \cite{dirfts_gama}. Seasonal feature drifts especially affect the building when less than a year of data is available for training \cite{LANGTRY2024114605, pinto_sharing_2022}. The main reason for this is that the training data does not yet include all seasons. However, seasonal drifts can also pose a problem after one year of data is available. For example, in an adaptive learning setup, the model may have overfitted on past seasonal data \cite{verwimp_continual_nodate, noauthor_overcoming_nodate}.

Other factors that strongly influence thermal dynamics modeling are \textbf{concept drifts}. Concept drifts refer to a change in the conditional distribution of the output given the input, typically caused by changes in the environment itself \cite{dirfts_gama, deng_2022_drift}. In buildings, such drifts can arise for various reasons. This study focuses on those most likely to occur and with the strongest impact on thermal dynamics and energy demand, namely retrofits and changes in occupancy \cite{haldi2011impact, wang2015methodology, bs2013_2322}. Another pronounced drift is a change in the heating system. This has no direct influence on the thermal dynamics of the building. It may only influence the supply temperature, making it a feature drift. This study, however, focuses on concept drifts. Seasonal drifts were included only because they occur naturally. 
%The proposed methodology is designed to be applicable to any household regardless of the heating system. Therefore, the system boundary for modelling is defined between the building and the heat source, making it independent of the specific heat source type. Hence, we did not consider this drift. 
A retrofit displays a concept drift, as it changes the building’s physical properties, typically by improving insulation. In contrast, changes in occupancy do not affect the building’s physical structure but influence its thermal behavior due to new heat gains and losses as well as different usage patterns. 
\edit{If occupancy is explicitly measured and included as an input feature in the prediction model, a change in occupancy can be interpreted as a feature drift. However, occupancy sensors do not provide all the necessary information, such as future window opening patterns or temperature set points. Also, occupancy sensors are rare in residential buildings and may not be adopted due to privacy concerns. Therefore,} occupant behavior is not included as input features in \edit{our} thermal dynamics model (see Section \ref{ch:pretrain}). As a result, the prediction model perceives changes in occupant behavior as drifts in the underlying system. Therefore, we classify changes in occupancy as concept drift. 

\begin{figure}[h!]
    \centering
    \includegraphics[width=1 \linewidth, trim={0.7cm 0 0.7cm 0},clip]{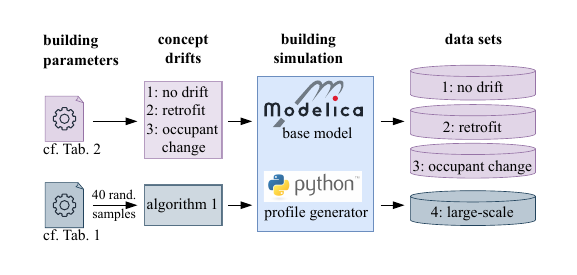}
    \vspace*{-0.7cm}
    \caption{Overview of generated datasets using the simulation framework according to \cite{builDaReview}. The first three datasets (purple) and the last one (gray) correspond to Section \ref{ch:concept_drifts} and Section \ref{ch:large-scale}, respectively.}
    \label{fig:data_generation}
\end{figure}

We simulate the 8 target buildings from Table \ref{tab:8targets} following the procedure in Figure \ref{fig:data_generation} (purple), considering three scenarios: (1) without drift, (2) with a retrofit, and (3) with a change in occupancy. Each target is simulated over five years with a 15-minute sampling interval. We select a five-year period to balance between observability of the concept drift and computational burdens for thermal dynamics modeling. 
%First, we generate one dataset without drifts, then a second dataset containing a retrofit, and a third containing a change in occupancy. This allows us to analyze the effect of each drift individually. 
The first dataset without concept drifts allows us to investigate the benefit of incoming unbiased data and the seasonal feature drifts (summer versus winter). The second and third datasets include a concept drift (retrofit and change in occupancy, respectively). We perform the drift for each target simultaneously to analyze the effect for modeling across several buildings. The drift occurs for all buildings on the first of April in the third year. This means that the transient process of modeling with little data in the starting phase has already subsided. This is especially crucial for a model from scratch, which needs several months to years of data for stable prediction performance. We apply the concept drift during spring to balance between summer and winter data, although the seasonal timing has no significant impact. After the drift, there are 33 months remaining to observe the transient process. Later, we will also investigate individual drift schedules in each target building (see Section \ref{ch:large-scale}).
%All targets experience the drift at the same time to clearly observe its effect.
%For the additional \textit{large-scale analysis}, we generate a larger data set for the period of seven years. 
For the retrofit (second dataset), we follow \cite{bottom_ibpsa_2017} and design the retrofit according to TABULA's building type L of single-family houses \cite{tabula}, which represents the building standard from 2016 until today. This results in a U-value of $0.11 [W/m^2K]$ for the outer walls and roof, as well as $0.7 [W/m^2K]$ for the windows. For the change in occupancy (third dataset), we employ a newly generated user profile for each target (see Section \ref{ch:data}). This affects temperature setpoints, internal gains, and window opening patterns. \editt{Table \ref{tab:drift_gen} summarizes the parameters affected by each drift.}

\begin{table}[h]
\centering
\begin{tabular}{m{0.4cm} | m{7.5cm}}
\rotatebox{90}{Change in occupancy} &
\begin{enumerate}[nosep, leftmargin=*, topsep=4pt]
    \item Select random number of residents
    \item Use number of residents to generate occupancy and window opening schedules according to \cite{builDaReview}.
    \item Temperature setpoint is selected between 20 and 24 \textdegree C, with an additional night setback between 0.5 and 4 \textdegree C for 70\% of profiles.
\end{enumerate}
\\ \hline
\rotatebox{90}{Retrofit  } &
\begin{enumerate}[nosep, leftmargin=*, topsep=4pt]
    \item Occupancy profile remains unchanged
    \item Set $U$-value$_{roof}$ = $U$-value$_{wall}$ = 0.11
    \item Set $U$-value$_{win}$ = 0.7
\end{enumerate}
\\
\end{tabular}
\caption{\editt{Summary of building parameter changes during drift occurrence.}}
\label{tab:drift_gen}
\end{table}

\begin{figure}[h!]
    \centering
    \includegraphics[width=1 \linewidth]{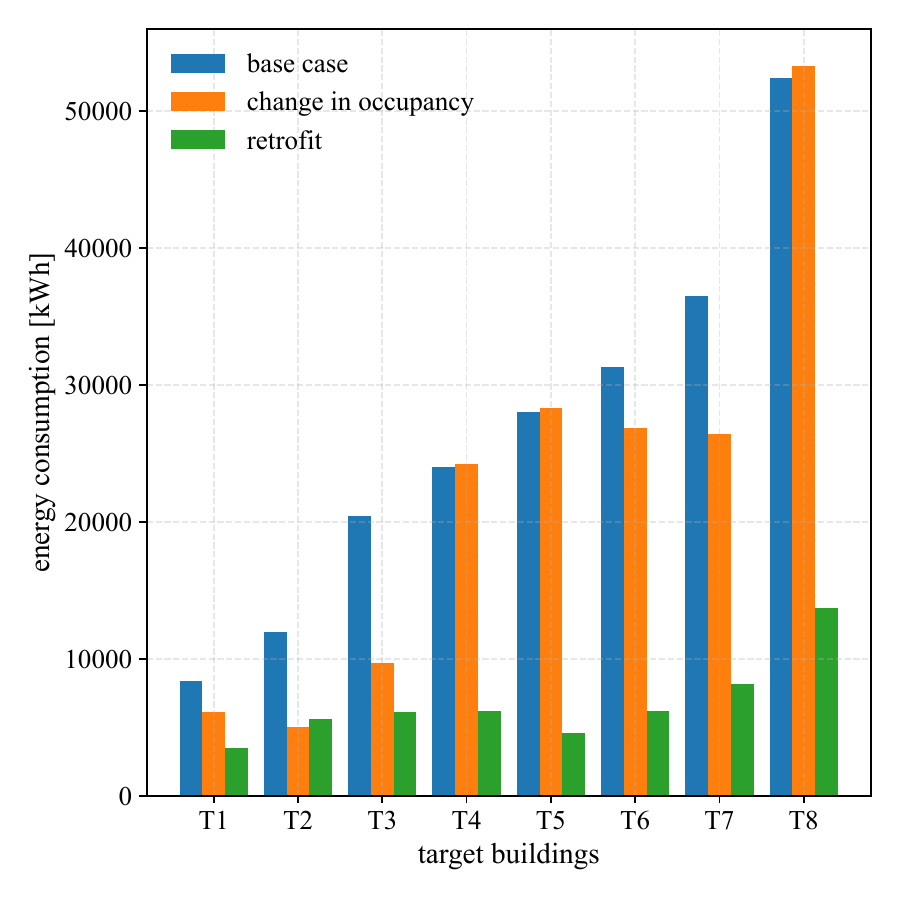}
    \vspace*{-1cm}
    \caption{Comparison of yearly energy consumption for the target buildings from Table \ref{tab:8targets} before (base case) and after the drift (change in occupancy \& retrofit).}
    \label{fig:energy_consumption}
\end{figure}

To demonstrate the effect of the concept drifts, we plot the yearly energy consumption before and after the drift in Figure \ref{fig:energy_consumption}. It shows that the change in occupancy may influence the energy consumption in a negative or positive way. In contrast, the retrofit always leads to a large decrease in energy demand.

%For the ventilation system, we consider an air change rate of $0.5 []$ and a heat recovery rate of $x []$.
%In Section \ref{ch:results} we will perform a \textit{small} and \textit{large scale} study. For the \textit{small scale} study all concepts drifts occur in the beginning of the third year. Thereby, the transient process in the beginning and effects from data collection are mostly omitted. Also, the effect of the concept drift can be analyzed more distinctly. 

\newpage
\subsection{Large-scale analysis}
\label{ch:large-scale}

For a more robust evaluation, we additionally perform a large-scale analysis. Therefore, we use 40 randomly drawn target buildings from the distribution of Table \ref{tab:targets_data} with individual drifting schedules to generate a fourth dataset, as illustrated in Figure \ref{fig:data_generation} (gray). %For the analysis of this dataset, we use a subset of more efficient methods for modeling. This allows us to consider a longer dataset, which we select to seven years. 
For this dataset, we choose a slightly longer period of seven years.
%For each target building, we develop an individual schedule for the concept drifts. 
Since the actual realization of a concept drift and its timing depend on the building owner or user, we develop an individual schedule for each target using a probability distribution for drift occurrence. This may include one retrofit and up to three changes in occupancy over the period of seven years. We set the chance for a retrofit to happen to 70\%, meaning for 30\% of the simulations, no retrofit occurs. Inspired by \cite{bernard2020does, sanchez2011move}, we determine the number of changes in occupancy as follows: 0 with 25\%, 1 with 35\%, 2 with 30\%, and 3 with 10\% probability. Once the number of concept drifts (retrofits and occupancy changes) is selected, we generate a schedule for each target by randomly assigning dates for each drift within the seven-year period. If, by chance, a user change and a retrofit were to happen within a month, the changes are applied simultaneously. Two occupancy changes are at least one month apart from each other.
Algorithm \ref{alg:data_gen} illustrates the overall schedule generation for the large-scale analysis, including retrofits and changes in occupancy for each target.
%TODO: How is the timeseries in the end gathered?

\begin{comment}
\begin{algorithm}[h]
    $parameters \gets parameters\_of(target)$ \; \
    $occupancy \gets rand\_occupancy()$ \; \
    $occ\_changes = weighted\_random\_nr(0, 4, prob\_occ)$\; \
    $ret\_changes = weighted\_random\_nr(0, 1, prob\_ret)$\; \
    $schedule = rand\_schedule(nr\_changes, ret\_changes)$\; \
    $start \gets 0$\; \
    \For{event, time in  schedule}{
        $new\_time\_series \gets simulate\_fmu(start, time, occupancy, parameters)$\; \
        $target\_time\_series.append(new\_time\_series)$ \; \
        \textbf{if} event ==1 : $occupancy \gets rand\_occupancy()$ \; \
        \textbf{if} event ==2 : $parameters \gets rand\_retrofit()$ \; \
        $start \gets time$\; 
    }
    \FuncSty{$weighted\_random\_nr(min, max, weights):$}{ \\
        \quad return random value from min to max based on weights \;
    }
    \FuncSty{$rand\_schedule(nr\_changes, nr\_change):$}{ \\
        \quad returns a list of tuples for seven years with \; \
        \quad 1: occupancy change\; \
        \quad 2: retrofit\; \
        \quad 3: end\; \
    }
    \caption{Data generation algorithm for one target building}
    \label{algo:data_gen}
\end{algorithm}
\end{comment}

\begin{algorithm}[tb]
   \caption{Data generation for one target building}
   \label{alg:data_gen}
   \begin{algorithmic}[1]
        \STATE \textbf{Given:} target building with random occupancy
        \STATE timeseries $\leftarrow$ [\:]
        \STATE start $\leftarrow$ 0
        \STATE schedule $\leftarrow$ generate\_random\_schedule()
        \FOR{event, time \textbf{in} schedule}
            \STATE snippet $\leftarrow$ simulate(from=start, to=time)
            \STATE timeseries = timeseries $\cup$ snippet
            \STATE \textbf{if} event is occ \textbf{then} randomize\_occupancy()
            \STATE \textbf{if} event is retro \textbf{then} do\_retrofit()
            \STATE \textbf{if} event is end \textbf{then} end\_simulation()
            \STATE start $\leftarrow$ time + 1
        \ENDFOR
        \STATE \textbf{return} timeseries
    \end{algorithmic}
\end{algorithm}

\subsection{Modeling building thermal dynamics}
\label{ch:pretrain}

%In this section, we discuss the general procedure of thermal dynamics modeling in buildings. This procedure usually consists of a training phase and an application phase, similar to the typical setup in Machine Learning. In this work, we consider this procedure as the initial phase of modeling, as depicted in Figure \ref{fig:setup} for the update step 1. 
%Either time-consuming modeling is recorded manually, or data-driven models are employed. Data-driven models offer the advantage of large-scale application. %continuous learning because they provide the basis for an update based on data. 
Building thermal dynamics have proven hard to model \cite{Drgona.2020}. This involves manual and time-intensive modeling, or data-driven approaches that enable a more scalable application across multiple buildings. However, data-driven models require several months to years of data to achieve good prediction performance. To address this, Transfer Learning (TL) has gained increasing attention recently \cite{peirelinck2022transfer, pinto2022transfer}. TL incorporates a pretrained model from a source building to model thermal dynamics of a target. Therefore, little data (weeks to months) from the target building is employed to fine-tune the pretrained model. This has the major advantage of reduced data requirements in the target building to achieve good prediction performance. Our study investigates the prediction performance of a target model from the time of implementation $t_0$ up to multiple years afterwards. To accelerate learning in the initial phase, we employ a TL model, which can be considered the state of the art in thermal dynamics modeling if a pretrained model is available \cite{raisch2025gentlgeneraltransferlearning, li_building_2024}. For comparison, we train ML models from scratch, which is the conventional approach when no TL is applied. A model from scratch trains a randomly initialized network. \edit{An alternative option for modeling building thermal dynamics is to employ generally pretrained time series models, like Chronos \cite{ansari2024chronos}, TimeGPT \cite{garza2023timegpt}, or TimesNet \cite{wu2022timesnet}. However, a comparison of their performance \cite{10.1145/3671127.3698177} on the Ecobee dataset \cite{ecobee} with that of TL models using the same dataset \cite{pinto_sharing_2022, li_building_2024} indicates that TL-based approaches and models trained from scratch achieve superior prediction accuracy. Moreover, edge devices within building management systems typically lack the on-site computational capacity required for retraining or fine-tuning such large pretrained general models. For these reasons, we focus on TL models and models trained from scratch.}

\begin{figure}[t]
    \centering
    \includegraphics[width=1 \linewidth]{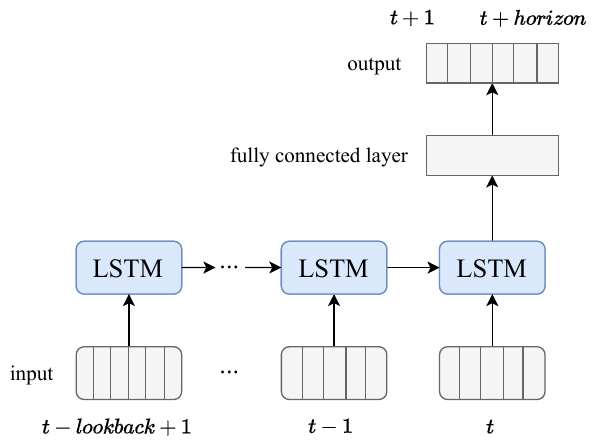}
    \caption{Architecture of the Neural Net.}
    \label{fig:architecture}
\end{figure}

The source selection procedure in TL is important but challenging \cite{li_building_2024}. Therefore, the authors of \cite{raisch2025gentlgeneraltransferlearning} demonstrated a general pretrained source model for TL. This model incorporated not only one source building but 450 different buildings. Thereby, the source selection process is omitted, as the general TL model can be used as a universal source for fine-tuning any target from the domain of single-family houses in Central Europe. The authors also reported lower prediction errors when using the general source model compared to both the conventional single-source TL approach and training from scratch. For these reasons, we reuse this model and employ it as a starting point for modeling our target buildings. Furthermore, we employ this model in the adaptive learning process, as described in Section \ref{ch:CL}. Throughout this work, we refer to it as the general TL model.
We inherit the architecture from the general TL model, which is a three-layer LSTM, as shown in Figure \ref{fig:architecture}. LSTMs were tested positively in their study and in others for building dynamics prediction \cite{li_building_2024, pinto_sharing_2022, Mtibaa2020LSTM, Martinez2020LSTMMLP}. \edit{Compared to more complex architectures such as Transformers, LSTMs offer a favorable balance between performance, training cost, and data requirements, which is advantageous in settings with limited computational resources and data availability, such as edge device operations in buildings. Their recurrent structure naturally captures the temporal evolution of state-space dynamic systems, making them well-suited for building thermal modeling and control tasks \cite{PILLONETTO2025111907}.}
The LSTM architecture includes a fully connected layer to generate an output sequence with the length of the forecast horizon. As a forecast horizon, we use one hour, similar to related literature \cite{raisch2025gentlgeneraltransferlearning, pinto_sharing_2022, chen_transfer_2020, li_building_2024}. \edit{With a 15-minute sampling time step, the output sequence results in four nodes in the fully connected layer.}
%Also this is a reasonable choice for MPC, since usually the control model is of the form $x_{k+1}=f(x_{k}, u_{k}, d_{k})$, with $x$ being the states, $u$ the control actions, $d$ the disturbances, and $f$ the thermal dynamics model \cite{Drgona.2020}.%For further information, please refer to their paper.
The task of the model is to predict the future indoor temperature (output) depending on current and past states (input). The considered states, respective inputs, are the room temperature $T_{in}$, the outside temperature $T_{out}$, the direct and diffuse solar irradiation $Q_{dir}$, and $Q_{dif}$, and the heat source control signal $u_{in}$, each from the current and past time steps. The amount of past time steps used for the input is called the lookback and is interpreted as a hyperparameter. We adopt the hyperparameters from \cite{raisch2025gentlgeneraltransferlearning}, as they performed hyperparameter tuning. \edit{The lookback is selected to 96, representing 24 hours of past time steps. Table \ref{tab:hyperparameter} summarizes the selected hyperparameters of the model.}

\begin{table}[!b]
    \centering
    \begin{tabular}{p{1.3cm} p{1.2cm} p{0.9cm} p{1.3cm} p{0.9cm} p{0.9cm} }
        \hline 
        \textbf{Parameter} & Lookback & LSTM layers & Neurons per layer & Output size & Input size \\ 
        \textbf{Selected}  & 96 & 3 & 125 & $4 \times 1$ & $96 \times 5$ \\ 
        \hline
    \end{tabular}
    \caption{\edit{Hyperparameter selection based on tuning according to \cite{raisch2025gentlgeneraltransferlearning}.}}
    \label{tab:hyperparameter}
\end{table}

For fine-tuning the general TL model, we use the available target data. Further discussion on the amount and availability of that data will follow in Section \ref{ch:CL}. We split the available target data into a train and a validation set with 70\% and 30\%, respectively. We utilize the training set for fine-tuning the parameters of the neural network. During the fine-tuning process, we use a strategy called best model selection to prevent overfitting on the training set \cite{Prechelt1998}. Best model selection evaluates the model’s performance on the validation set over training epochs and selects the model that achieves the lowest validation error. 
For training models from scratch and for the model update (see Section \ref{ch:CL}), we use the same train-validation split.
%For training models from scratch, or for updating models over time (cf. Section \ref{ch:CL}), we follow the same train-validation split approach.

%In this section, we discussed the general procedure for thermal dynamics modeling in buildings. This procedure follows the typical setup in Machine Learning, consisting of a training phase and an application phase. 
%However, this approach may be impractical for modeling buildings for two reasons. First, new measurement data is continuously collected and can be used to improve the model. Second, buildings are often subject to changing environmental conditions, which highlights the need for an adaptive model. 

\subsection{Updating dynamics model during operation}
\label{ch:CL}

%initial phase of modeling, which employs either the general TL model or a model trained from scratch. However, during building operation, new operational data will be measured. The new measurement data can be used to update the initial model and thereby adapt the model to the current building state and improve its performance. We denote this process as \textbf{adaptive learning}. To investigate adaptive learning, we follow the setup according to Figure \ref{fig:setup}.
In Section \ref{ch:pretrain}, we discussed the general procedure for modeling thermal dynamics. This procedure follows the typical setup in Machine Learning, consisting of an initial training or fine-tuning phase and a subsequent application phase. This structure applies to both TL and modeling from scratch \cite{raisch2025gentlgeneraltransferlearning, li_building_2024, Mtibaa2020LSTM, Martinez2020LSTMMLP}.
However, this might be impractical for modeling buildings for two reasons. First, new measurement data are continuously collected and can be used to improve model performance. Second, buildings are subject to changing environmental conditions (see Section \ref{ch:concept_drifts}), which highlights the need for an adaptive model. 
For these reasons, we continue with \textbf{adaptive learning} after the initial fine-tuning phase. Adaptive learning refers to the continuous updating of a model during operation. 
%This paper focuses on identifying the best strategy for updating the thermal dynamics model with regard to 
For the model update, we consider methods from the domains Transfer Learning (TL), Continual Learning (CL), and training from scratch.
%Different strategies can be employed to update the model. It is the task of this paper to find the best strategy for thermal dynamics modeling with regard to drifting and non-drifting building environments. 
The general process of adaptive learning is illustrated in Figure \ref{fig:setup}. Figure \ref{fig:setup} (a) displays the data accumulation and the updating steps over time, and Figure \ref{fig:setup} (b) displays the model update. 

The modeling of a particular target building starts with the initial fine-tuning phase at time instance $t_0$. During period $x_{1}$, we collect the first target data. This data is employed to train the initial model by fine-tuning the general TL model (see Section \ref{ch:pretrain}). 
After the initial fine-tuning phase, we proceed in time and update the model with a particular adaptive learning strategy. The model update appears in an iterative manner after each update period $|x|$, similar to the approaches in \cite{li_large-scale_2023, zhou_elastic_2022, pinto_sharing_2022}. For each model update, the newly collected data $x_{n}$ as well as the old data $\{x_{1},\dots,x_{n-1}\}$ can be used. Updating methods follow different strategies regarding which and how much data they use. In the following, we will distinguish between strategies based on adaptive TL (Domain: TL) and Continual Learning (Domain: CL). For adaptive TL, we will employ the general TL model $G(.)$ in every update step, as illustrated in Figure \ref{fig:setup} (b). For CL, we will fine-tune the specific target model from the previous time step ($f_{n-1}(.)$ for $n\geq2$). In any case, the base model ($G(.)$ or $f_{n-1}(.)$) serves as weight initialization, meaning all parameters of the pretrained model are adjusted during the update. 
From the domain of CL, various methods can be employed for the model update. We follow \cite{li_large-scale_2023} and \cite{zhou_elastic_2022} and employ their best-performing methods, since they applied CL to a similar task. 
%While many other approaches to adaptive learning exist, this study can only evaluate a selected subset of promising methods due to computational constraints. 
Below, we briefly explain each considered method from the domains TL, CL, and the model from scratch. A summary can be found in Table \ref{tab:cl_algorithms}:

%\subsubsection*{Continual Backpropagation}
%First, we consider the Continual Backpropagation Algorithm (CBA), according to \cite{dohare_loss_2024}. This algorithm displays the latest contribution in CL and illustrates good performance in the ablations shown by the authors. Their strategy randomly reinitializes neurons within the network based on a probability value. The idea behind this is to reactivate dormant units of the network. However, this strategy was only shown for a feedforward neural network. Therefore, we are the first to introduce this strategy for LSTMs.
%\textbf{Explain differences in approaches of CL methods - i.e. how technical approaches are they taking} E.g. gradient-based vs loss term vs reinitialisation to escape local min.

\subsubsection*{Incremental Learning (Domain: CL)}
Incremental Learning (IL) uses only the newly collected data for updating the model from the previous time step. No data from previous update steps is used\footnote{We also tested incorporating both new and previously observed data to update the target model. This approach, known as Accumulative Learning (AL), provided no performance improvement or additional insights but required high computational resources (O(N) memory complexity). Therefore, AL was excluded from the results.}. 
Thereby, IL displays the simplest CL method. Despite this, it was found to be best for gradual drifts in electric load forecasting \cite{li_large-scale_2023}.

\subsubsection*{Elastic Weight Consolidation (Domain: CL)}
Elastic Weight Consolidation (EWC) \cite{noauthor_overcoming_nodate} reduces catastrophic forgetting by identifying important weights from previous updates and penalizing unfavorable changes to them. It estimates the importance of each parameter using the Fisher information matrix and adds a regularization term to the loss function accordingly. This term discourages updates that would alter critical parameters, allowing the model to retain performance from earlier updates while learning on new data. In our implementation, EWC uses a buffer of 1000 training examples, which is updated with 250 new examples at each iteration.
%Our implementation implicitly uses examples from more recent update steps more frequently for the regularization, penalizing negative changes that affect more recent dynamics more.

\subsubsection*{Gradient Episodic Memory (Domain: CL)}
Gradient Episodic Memory (GEM) \cite{lopez-paz_gradient_2022} stores a subset of past data in episodic memory, in our case 250 random samples from each previous update. GEM constrains gradient updates during training to prevent an increase in loss on the stored samples. This prevents the model from forgetting previously learned dynamics by ensuring compatibility with previous knowledge.
Unlike EWC, which relies on parameter importance and regularization, GEM directly constrains the gradient updates.

\subsubsection*{Seasonal Memory Learning (Domain: CL)}
During the experiments, we observed variations in performance across seasons. To address this issue, we propose incorporating data from both the current update period and the corresponding period in the previous year. However, this is only applicable once more than one year of data from the target building is available. For the n\textsubscript{th} update we use the data $x_n$ plus $x_{n-\alpha-1}, x_{n-\alpha}, x_{n-\alpha+1}$ with $\alpha = 12 /|x|$ and $|x|$ being the update-period in months. We use this data to update the model from the previous update step. In the following, we call this method Seasonal Memory Learning (SML).

%\subsubsection*{Accumulative Learning}
%Last, the most straightforward case is also considered, which is Accumulative Learning (AL). All collected data is used for AL to train a Scratch model.\\
\subsubsection*{Incremental Learning on General Model (Domain: TL)}
Methods of the domain TL fine-tune the general TL model in each update step. Incremental Learning on General Model (GIL) employs only the new data for the update. This results in a model that is specifically fine-tuned to the target building and its current update step.

\subsubsection*{Accumulative Learning on General Model (Domain: TL)}
Accumulative Learning on General Model (ALG) uses all collected data in the target (past and new data) to fine-tune the general model. This illustrates a similar approach to a model from scratch for each update step, as all available data is used for training. However, the difference is that the general TL model serves as the weight initialization of the neural network. We also propose a variant of ALG, called event-based Accumulative Learning on General Model (eALG). This method discards all pre-event data and uses only post-event data for subsequent updates. Such an event can be a concept drift, for example.

\subsubsection*{Initial fine-tuning (Domain: TL)}
Initial fine-tuning (IFT) means the conventional case as used in TL literature \cite{raisch2025gentlgeneraltransferlearning, pinto_sharing_2022, li_building_2024} when no adaptive learning is applied. The pretrained model (in our case, the general TL model) is fine-tuned once on a fixed set of the target building's data, and no update is performed over time. This approach is similar to the standard ML model that is trained once and applied thereafter. The amount of data used for IFT is a design choice and consists of the assumption on how much data is necessary to receive sufficient performance. As we assess multiple update periods $|x|$, we also account for several initial fine-tuning periods in the experiments in Section \ref{ch:results}.
In the following, we consider IFT as the benchmark to compare the performance of the adaptive learning methods. This provides information on how much better adaptive learning is compared to not updating models, meaning the standard ML approach. A model trained from scratch on the initial training data that is not updated over time could be an alternative benchmark. However, IFT outperformed models from scratch and is therefore preferred \cite{raisch2025gentlgeneraltransferlearning}.

\subsubsection*{Adaptive Model from Scratch (Domain: Scratch)}
Training a model from scratch is the conventional way when no CL or TL strategies are employed (cf. Section \ref{ch:pretrain}). A model from scratch means to train a model with randomly initialized weights, as illustrated in Figure \ref{fig:setup} (b). In an adaptive setting, a new model is trained for each update period n, using all accumulated data over time \cite{pinto_sharing_2022}. This is the only strategy that does not include the general TL model in the initial modeling phase.

\begin{table*}[t!]
\caption{Comparison of update strategies by data requirements and memory complexity (Big-O notation). Methods having domain CL employ the model from the previous update step $f_{n-1}$ for weight initialization. Methods having domain TL use the general TL model as weight initialization. N represents the number of updates.}
\vspace{0.1cm}
\label{tab:cl_algorithms}
\centering
\renewcommand{\arraystretch}{1.2}
\begin{tabular}{c p{5.9cm} c p{5cm} c} 
\hline
\multicolumn{2}{l}{\hspace{1.5em}\textbf{Method}} & \textbf{Domain} & \textbf{Data Requirements}  & \textbf{Memory Complexity}\\ 
\hline
Scratch & Adaptive Model from Scratch 
& Scratch & new + old data & O(N) \\

IL & Incremental Learning 
& CL & new data & O(1) \\

EWC & Elastic Weight Consolidation \cite{noauthor_overcoming_nodate} 
& CL & new data + 1000 examples representing previous updates & O(1) \\

GEM & Gradient Episodic Memory \cite{lopez-paz_gradient_2022} 
& CL & new data + 250 random examples from each previous update & O(N) \\

SML & Seasonal Memory Learning 
& CL & new data + data from previous year but same season & O(1) \\

GIL & Incremental Learning on General Model 
& TL & new data & O(1) \\

ALG & Accumulative Learning on General Model 
& TL & new + old data & O(N) \\

eALG & Event-based Accumulative Learning on General Model 
& TL & new data + old data after last event & O(N) \\

IFT & Initial fine-tuning 
& TL & only $x_1$ period of data & O(1)  \\
\hline
\end{tabular}
\vspace*{0.5cm}
\end{table*}

%The computational costs are the time required to train the model. The memory costs are the amount of memory required to store the necessary data for each method. Table \ref{tab:cl_algorithms} shows a comparison regarding the memory costs for all methods using the Big-O notation. A method has memory complexity O(1) if it uses a constant amount of data independent of the update step, and O(N) if its data usage grows proportionally with the update step.
%CL focuses in general on using the least amount of memory and computational (training) time while maximizing model performance. TL and a model from scratch do not aim for the tradeoff but rather focus on model performance. 

\subsection{Evaluation}
\label{ch:evaluation}

%In this study, we want to evaluate the effect of concept drifts on the prediction performance for thermal dynamics modeling. Therefore, we analyze the evolution of the test error over time during building operation. For the upcoming experiments, we will either report the error over time or the overall error over the considered period.
%Specifically, we plot the test error at each update step, continuing until a maximum duration of five years is reached. Since multiple buildings are considered, we report the mean and variance of the test error across all target buildings for each update step. In addition, we present the mean test error over the entire five-year period, the annual mean error, and the corresponding variances.
We evaluate the updated model $f_n (.)$ using the test set, shown in blue in Figure \ref{fig:setup}. The test set represents the unseen data in the consecutive order of the current update period $x_n$. This is the period the new model would operate on in a real-world application. As the model is only updated after a full updating period, the testing period is set to match this duration. For the upcoming experiments, we analyze the evolution of the test error for each proposed method over time during building operation. 

To evaluate the performance, we use \edit{five} error metrics: \edit{the mean absolute error (MAE), the coefficient of determination (R²)}, the root mean square error (RMSE), the mean absolute scaled error (MASE), and the relative RMSE improvement (RRI):

\newcommand{\eqnsep}{2ex}
\begin{align}
    \text{MAE} &= \frac{1}{n} \sum_{i=1}^{n} \frac{1}{h} \sum_{j=1}^{h} |T_{in,i+j} - \hat{T}_{in,i+j}| \label{eq:mae} \\[\eqnsep]
    \text{R}^2 &= 1 - \frac{\sum_{i=1}^{n} \sum_{j=1}^{h} (T_{in,i+j} - \hat{T}_{in,i+j})^2}{\sum_{i=1}^{n} \sum_{j=1}^{h} (T_{in,i+j} - \bar{T}_{in})^2} \label{eq:r2} \\[\eqnsep]
    \text{RMSE} &= \sqrt{\frac{1}{n} \sum_{i=1}^{n} \frac{1}{h} \sum_{j=1}^{h} (T_{in,i+j} - \hat{T}_{in,i+j})^2} \label{eq:rmse} \\[\eqnsep]
    \text{MASE} &= \frac{\frac{1}{n} \sum_{i=1}^{n} \frac{1}{h} \sum_{j=1}^{h} |T_{in,i+j} - \hat{T}_{in,i+j}|}    
    {\frac{1}{n} \sum_{i=1}^{n} \frac{1}{h} \sum_{j=1}^{h} |T_{in,i+j} - T_{in,i}|} \label{eq:mase} \\[\eqnsep]
    \text{RRI} &= \frac{\text{RMSE}_{\text{Benchmark}}- \text{RMSE}_{\text{Model}}}{\text{RMSE}_{\text{Benchmark}}} \label{eq:RRI}
\end{align}

where n is the number of training examples, h the horizon, $T_{in}$ the true temperature value, \edit{$\bar{T}_{in}$ the mean of the true temperatures}, and $\hat{T}_{in}$ the predicted value.
 %The MAE reports the average absolute error without giving extra weight to outliers. 
\edit{The MAE reports the average absolute difference between the predicted and observed temperature in degrees Celsius. The R² indicates how well the model explains the variance in the data, with values close to 1 denoting a better fit.}
The RMSE penalizes large errors more strongly \edit{compared to MAE}.
The MASE is a scaled metric that compares a model's performance to a naive predictor. The naive predictor uses the last observed output as a constant forecast over the entire horizon. Thereby, the metric provides some insight into the difficulty of the prediction task. An MASE value below 1 indicates that the model outperforms the naive predictor.
The RRI provides a relative improvement regarding a benchmarking method. We consider initial fine-tuning (IFT) as the benchmark, as explained in Section \ref{ch:CL}. 
%This method does not incorporate incoming data over time. Therefore, it displays the standard ML case \cite{verwimp_continual_nodate, CL_review} as well as the classic method of related literature \cite{raisch2025gentlgeneraltransferlearning, li_building_2024, CHEN20191141}.

For model training, we used an AIME T600 workstation\footnote{With NVIDIA RTX A6000 GPU (48 GB), Threadripper Pro 5995WX CPU (64 cores, 2.7 / 4.5 GHz), 512 GB RAM.}. The methods allocated around 2--3 GB of GPU memory. We observed differences in computational and memory costs for each method. A detailed discussion of the computational requirements is provided in Section \ref{ch:compute}. The code for all of the proposed methods is available on GitHub \cite{CLGit}.

%\newpage
\section{Experiments}
\label{ch:results}

In the following, we present the experiments to address the previously stated research questions.
%Depending on the context, we either report the error trajectory over time or the aggregated error over the full evaluation period.

Section \ref{ch:ex_undrifted} investigates the benefit of incorporating new data when no concept drift occurs. For this purpose, we use dataset 1 from Section \ref{ch:concept_drifts}. This serves as the baseline study. Additionally, we include a sensitivity study for the update period and a more granular perspective to analyze seasonal drifts. 
%Section \ref{ch:ex_undrifted} displays the baseline study that employs data that is not affected by concept drifts. For this purpose, we use dataset 1 from Section \ref{ch:concept_drifts} without changes. Additionally, we include a sensitivity study for the update period and a more granular perspective to analyze for seasonal drifts. 

After this baseline study, we analyze the effects of concept drifts for adaptive learning. Section \ref{ch:ex_concept_drifts} examines buildings affected by retrofits and changes in occupancy. For these studies, we use datasets 2 and 3 as described in Section \ref{ch:concept_drifts}. 

Finally, Section \ref{ch:ex_large} validates the results of the previous experiments by performing a large-scale analysis. This study uses five times more buildings and extends the evaluation period to seven years, as explained in Section \ref{ch:large-scale}.

\newpage
\subsection{Undrifted data}
\label{ch:ex_undrifted}

\begin{figure*}[!htp]
    \centering
    \includegraphics[width=\linewidth]{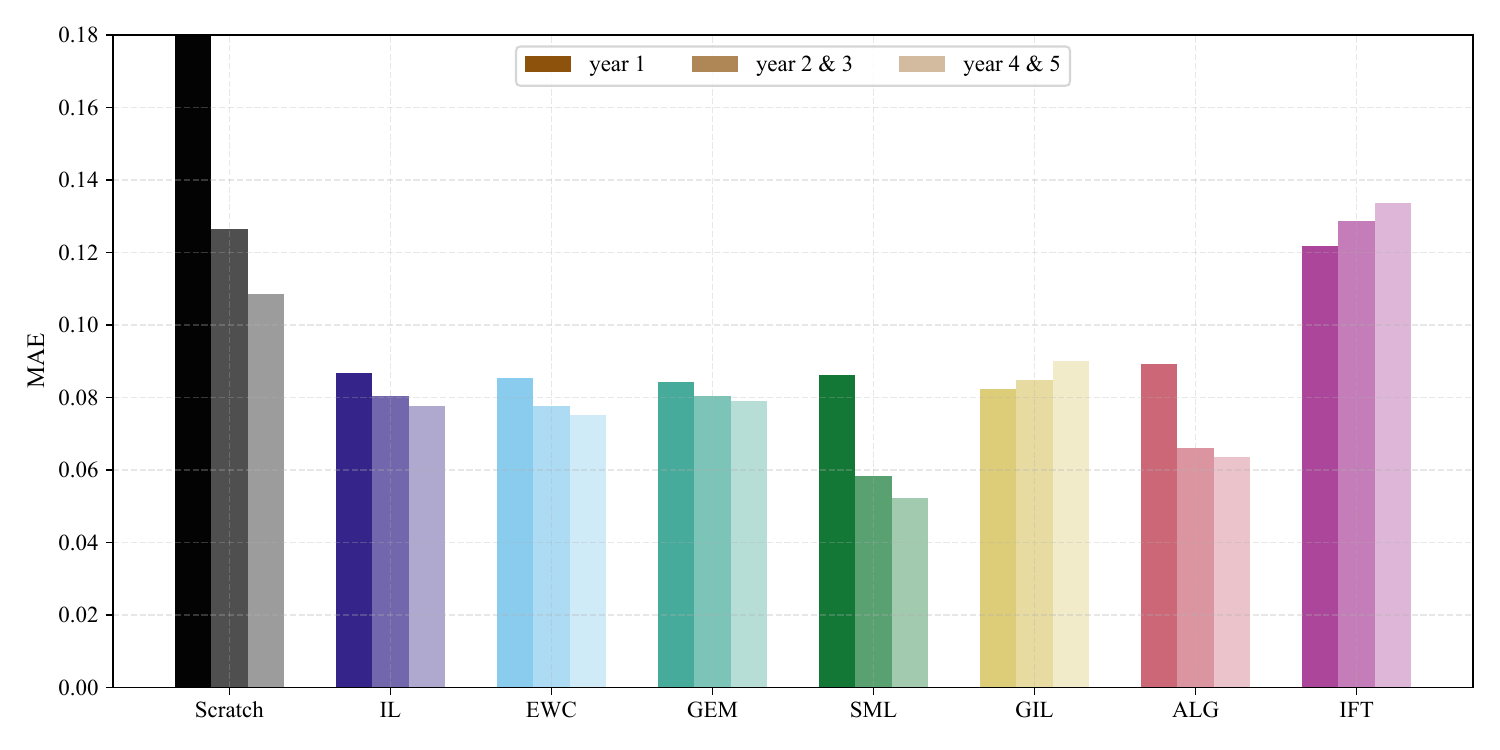}
    \vspace*{-1cm}
    \caption{Averaged \edit{MAE} values across all target buildings for different updating methods with a model updating period of one month and no concept drifts in the building. CL methods in green/blue and TL methods in yellow/red. The model from scratch exceeds bounds for the first year with an \edit{MAE of 0.519}.}
    \label{fig:bar_plot_yearly}
\end{figure*}

We first analyze how incorporating measurement data over time affects the accuracy of thermal dynamics models using undrifted data. This serves as the baseline study, representing buildings where both the fabric and occupant behavior remain unchanged during operation. All updating methods from Section \ref{ch:CL} are compared to the benchmark IFT, and their accuracy is observed over time. As a base case, we consider an updating period of one month, similar to \cite{raisch2025gentlgeneraltransferlearning, li_large-scale_2023}. Figure \ref{fig:bar_plot_yearly} illustrates the results as an average error across the 8 target buildings. For illustrative purposes, we only plot the error across three periods. We average the error across the period given in the legend. For detailed yearly results, we refer to Appendix \ref{tab:yearly_rmse}.
In the first year, we observe similar performance for all CL and TL strategies. The benchmark IFT, however, has a higher error, since it was only fine-tuned once within the first year, versus the other methods that had 12 updates. The model from scratch has a much higher prediction error in the first year due to the absence of a pretrained model.
The strategies that mainly focus on the new data, namely IL, EWC, GEM, and GIL (see Table \ref{tab:cl_algorithms}), seem to have a small performance increase over time. On the other hand, the methods that incorporate all available data, namely ALG and the model from scratch, have a large performance increase over time. SML incorporates new data and data from the previous year during the same season, leading to a similar performance increase to ALG. 
In general, the trend increase flattens after the 2nd and 3rd year.
The errors of IFT (benchmark) and GIL remain more or less the same over time. IFT does not perform any updates. GIL fine-tunes the general TL model at each update step, but only using new data. However, prediction performance varies over the years, most likely due to changes in weather conditions. The overall best-performing methods after five years are Accumulative Learning on General Model (ALG) and Seasonal Memory Learning (SML). \editt{Table \ref{tab:noise} in the Appendix also summarizes the results for data affected by artificial measurement noise. These results show an additional error corresponding to the introduced noise, which does not affect the relative performance of the different updating methods. Given that measurement noise varies considerably with sensor quality, and to improve the clarity of results, we report the remaining findings without additional noise.}

\begin{figure*}[!htp]
    \vspace*{-2\baselineskip}
    \centering
    \includegraphics[trim=0.3cm 0 0.1cm 0, clip, width=1 \linewidth]{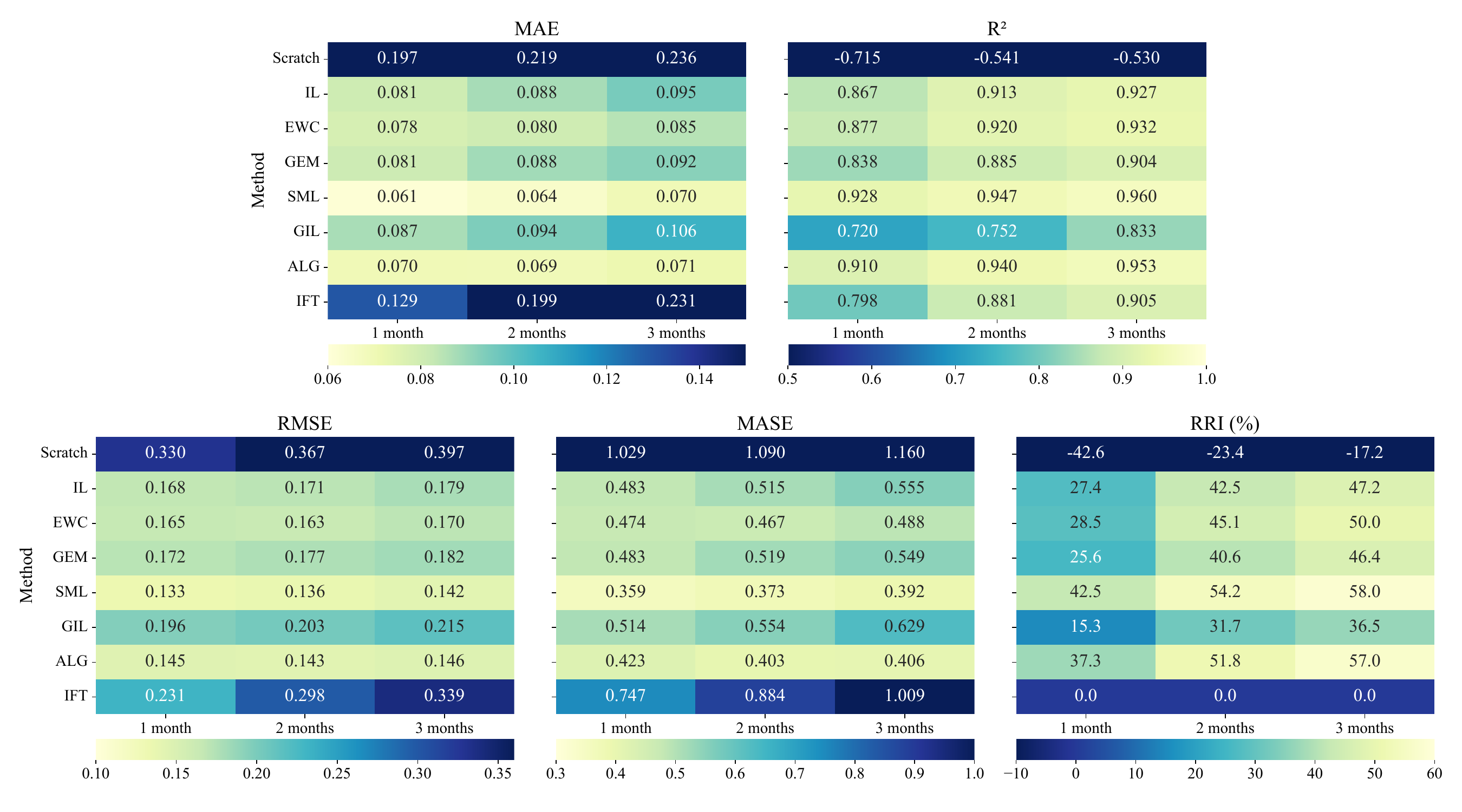}
    \vspace*{-1cm}
    \caption{Heatmap of prediction errors (\edit{MAE, R²,} RMSE, MASE, RRI) averaged across all target buildings over five years for different update periods.}
    \label{fig:heat_map}
\end{figure*}

The update period is a key factor in adaptive learning. It affects prediction accuracy, data used for each update, the computational cost, and the time required for the model to adapt to new environmental conditions \cite{verwimp_continual_nodate, li_large-scale_2023}. Hence, we include a sensitivity study regarding the updating period. We consider a one-month update period as the base case. A smaller amount of data would lead to unreasonable models due to the test-validation split and the resulting small validation set. Additionally, we investigate update periods of two and three months. Longer periods are unreasonable, as results will show, and because of the seasonal variations. For the analysis of the update period, we assess the averaged error from initial implementation (see Figure \ref{fig:setup} (a) $x_{1}$) until the end of the 5th year of operation across the 8 target buildings.
We demonstrate the results as a heatmap for the \edit{MAE, R²,} RMSE, MASE, and RRI values in Figure \ref{fig:heat_map}. Better performance is displayed in yellow, and worse in blue. Appendix \ref{tab:grouped_avg_errors} also summarizes the results in a table including the 95\% confidence intervals.
In general, we see the best performance regarding \edit{MAE} values for the updating period of one month. An update period of two months yields slightly worse results. \edit{For the MASE and RMSE values, EWC and ALG perform best with a two-month update, while the remaining ones perform best with a one-month update. The three-month update achieves the worst performance. R² performs best for a 3-month update period. However, this can be misleading, as the denominator of R² becomes larger with a longer testing period due to increased data variance resulting from seasonal changes. Therefore, an R² value closer to one for a 3-month update period does not necessarily indicate a better prediction model.
%The nominator is proportional to the sum of squared errors, increasing only slightly as indicated by MAE and RMSE values. 
The model trained from scratch exhibits a negative average R² due to large errors in the initial update steps. As more data become available, R² values improve and approach 1. The remaining methods consistently achieve higher average performance compared to the model from scratch.
}
%The initial fine-tuning period seems to have little influence on the IFT performance over the years. Only the MASE values decrease noticeably, meaning the IFT model performs better compared to the naive predictor with larger initial fine-tuning.
Across all updating periods \edit{and metrics}, the best strategy is \edit{SML (Domain: CL)}, closely followed by \edit{ALG (Domain: TL)}. Both with an average improvement of around \edit{40}\% compared to IFT (update period: one month). With regard to the standard CL methods, EWC performs the best.

In the following, we will use these results as a guide and focus on the most promising methods. These include the benchmark IFT, the best-performing CL method EWC, the overall best methods SML and ALG, as well as the model trained from scratch. The best update period was one month, closely followed by two months. 
%For scenarios with attention to best performance, the one-month update is used, and for a trade-off between performance and computational time, the update period is two months.
For best performance, a one-month update is used and for a trade-off between performance and computational time, the update period is two months.

%\editt{To describe the influence of building‐insulation quality on prediction performance, we sort the previous results by U-values. Figure \ref{fig:u_values_scatter} shows the results. The model trained from scratch performs worse for well-insulated buildings. IFT exhibits a U-shaped pattern, achieving its best performance at medium insulation levels. The updating-based methods show minor sensitivity to insulation quality, but they generally follow the same trend as the model trained from scratch. It should be noted that only two buildings fall into each U-value category, which limits the reliability of these observations. For more comprehensive studies, we refer to \cite{builDaReview, pinto2022sharing, builda2}.}

\begin{figure}[!p]
    \centering
    \includegraphics[trim=0.4cm 0 0.4cm 0, clip, width=\linewidth]{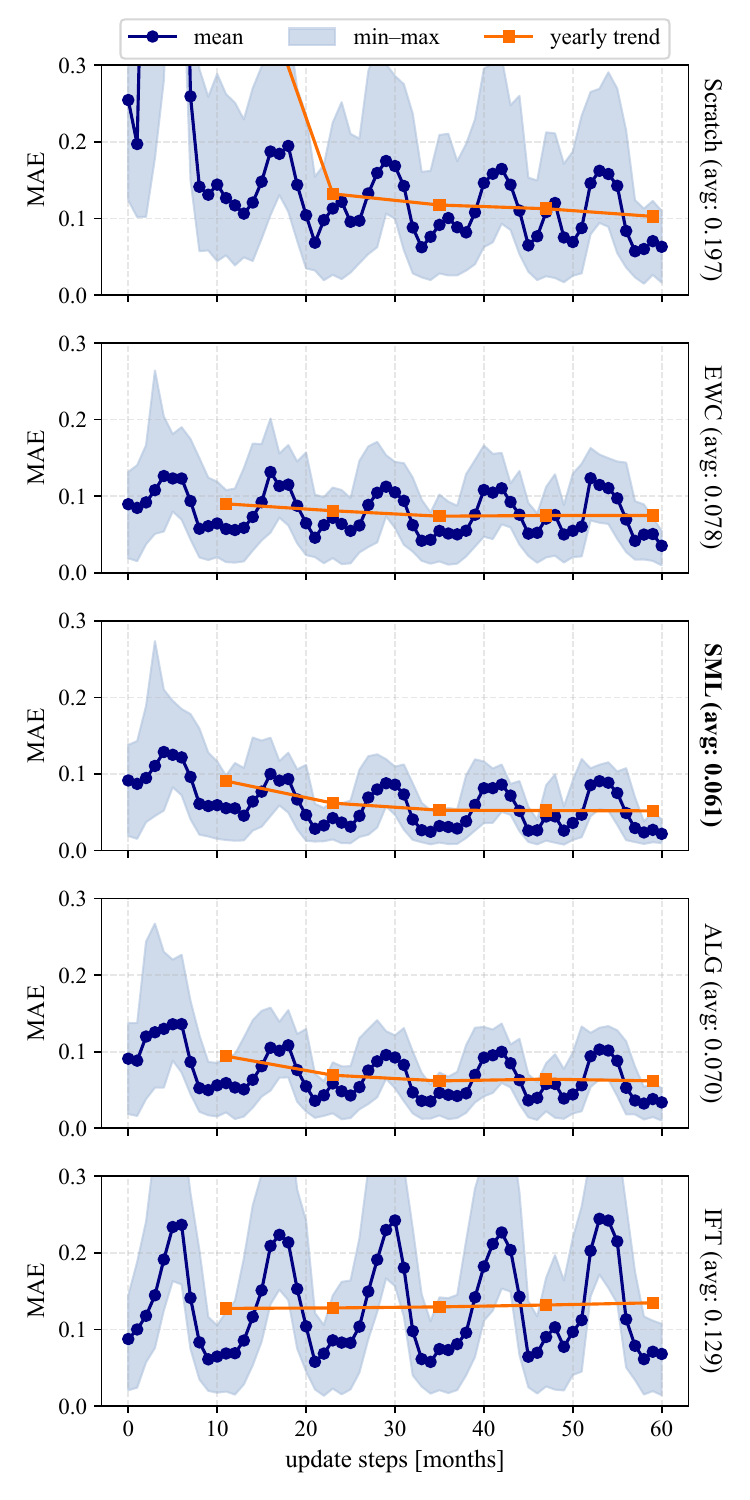}
    \vspace*{-1cm}
    \caption{\edit{MAE} values for each update test period (updating interval: one month) across all target buildings with no concept drifts in the environment.}
    \label{fig:noChange_over_time}
\end{figure}

Seasonal variation may affect thermal dynamics modeling, as discussed in Section \ref{ch:concept_drifts}. To analyze the effect of seasonality during adaptive learning, we use data unaffected by concept drifts. Drifted data could distort outcomes. We provide results at a granular level, i.e., for each update step (one month), to illustrate the seasonal performance over time.
%Additionally, we analyze seasonal drifts, i.e., performance variations between different periods of the year. For this purpose, we use the same dataset as in the baseline case, comprising five winter and five summer periods without concept drifts.We report the results for each update period (one month) to provide an understanding of the seasonal performance over time. 
Figure \ref{fig:noChange_over_time} displays the test \edit{MAE} averaged over the 8 target buildings, with the variance indicated by the min-max range. 
All methods show some seasonal behavior. Winter occurs in the beginning (the first testing period is in February) and then every 12 updating steps thereafter. The summer takes place in the opposite time intervals. Prediction accuracy is substantially better in the winter than in the summer. An explanation for this is that the indoor temperature in the winter is mostly driven by the heat control and the temperature setpoint. The indoor temperature typically does not fluctuate much around that setpoint. However, in the summer, the outside weather conditions drive the building, as it operates in free-floating conditions. This leads to larger temperature deviations and, as a result, to a more difficult prediction task.
In contrast to the benchmark IFT, all other methods display a trend of increasing performance as they follow an updating strategy. 
%This improvement is most evident in summer, while winter errors remain largely stable. 
The trend is visible during the first three years. The fifth year is more difficult to predict, likely due to specific weather conditions.
The variance of IFT remains constant over time. For EWC, the variance decreases slightly, following the performance trend. In contrast, SML, ALG, and the model trained from scratch show a stronger reduction in variance. After five years, SML and ALG achieve the lowest overall variance among all methods.

%The trend is most pronounced for the model trained from scratch. This is because it starts training from randomly initialized weights. Thereafter are the methods SML and ALG, which have a similar trend line. Also, EWC's impact on the trend is similar, though it is slightly lower. 

%\newpage
\subsection{Concept drifts}
\label{ch:ex_concept_drifts}

\begin{figure*}[p]
    \vspace*{-2\baselineskip}
    \centering
    \includegraphics[width=1 \linewidth]{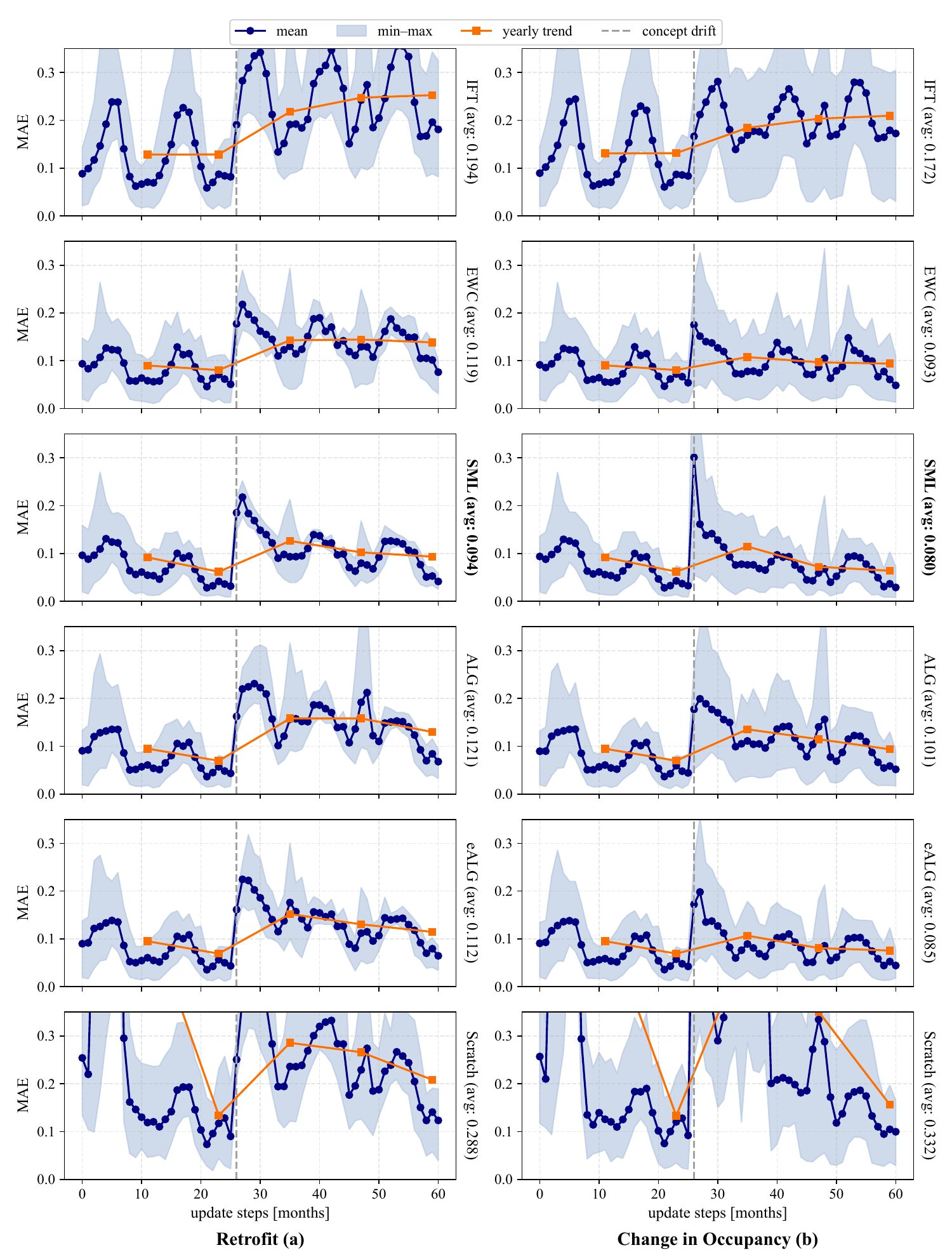}
    \vspace*{-1cm}
    \caption{\edit{MAE} values over time with a model updating period of one month for a retrofit and a change in occupancy after 2.25 years (26th update step).}
    \label{fig:combined_plot}
\end{figure*}

In the following, we analyze the effect of concept drifts for the adaptive learning setting, starting with a retrofit. The retrofit applies to all buildings on April 1 in the third year (26th update step).
Figure \ref{fig:combined_plot} (a) shows the test error across the 8 target buildings for each update step. All considered methods respond to the drift with an increase in the prediction error. 
The model from scratch reacts the strongest to the change. One main reason might be the absence of the general TL model, which generalizes better across different insulation levels. Another reason is the mix of drifted and undrifted data and the split of the training and validation set (see Section \ref{ch:pretrain}). Due to the split, it takes some time for the training set to include drifted data. 
%In contrast, the validation set, which is used for best model selection, already contains drifted data. 
This effect also applies to ALG, as the amount of data and the split are the same. Therefore, it is a reasonable option to neglect the previous data and only use data after the concept drift to fine-tune the general TL model. We call this option event-based Accumulative Learning on General Model (eALG), as explained in Section \ref{ch:CL}. For eALG, we observe improved performance. However, its overall behavior is similar to ALG, since it requires time to accumulate new data after the event. Furthermore, the error is larger for the drifted year compared to the first year (update 0 to 11). This indicates reduced performance of the general TL model for well-insulated buildings, since all considered targets correspond to this class after the retrofit. 
%This probably results from the training phase of the general TL model, where well-insulated buildings represent the edge case of the considered pertaining distribution. 
Nevertheless, we observe better results for all models that originate from the general TL model compared to the model from scratch. 
%This indicates that the model from scratch overfitted to the undrifted target building. In comparison, the other models seem to profit from the general TL model.
The benchmark IFT is most affected among the ones that used the general TL model for fine-tuning. Since IFT does not perform any updates, the error does not decrease after the drift.
The overall best-performing strategy is SML.

%The methods that do not incorporate old data, namely EWC and eALG have better results right after the drift than the methods that use old data. SML uses old seasonal data and new data. This results in a large error increase, but after that in a strong error decrease. 
\hfill\\
%We chose the updating period of two months since this yields a tradeoff between performance and computational resources. Section \ref{} provides a more in-depth discussion on computation.
Next, we examine the impact of changes in occupancy on model performance. Similar to the retrofit, the change in occupancy applies to all buildings at the same time, as explained in Section \ref{ch:concept_drifts}. Each target building receives its own new occupancy profile.
Figure \ref{fig:combined_plot} (b) illustrates the results. The effect is most pronounced for the model from scratch. After the drift, the average error and the variance significantly increase, due to the combination of drifted and undrifted data, as previously noted for the retrofit.
This effect is not as significant for ALG, although the same data and split are used. The reason for this might be the generalization capabilities of the general TL model to several user patterns. The change in occupancy affects ALG, eALG, EWC, and SML similarly. The average error and variance increase significantly after the drift, when the test data contains drifted data but the training data does not. In the following update steps, with drifted data included in the training set, the average error and variance stabilize again.
In contrast, the occupancy change affects the trend error and the variance of the benchmark IFT over long time as no updates are performed. However, the effect is not as significant as for the model from scratch, since IFT benefits from the general TL model.

\newpage
\subsection{Large-scale analysis}
\label{ch:ex_large}

In this section, we display the results for the large-scale analysis. This study aims to provide more robust results by covering a larger number of buildings with diverse drift profiles over a seven-year period, as described in Section \ref{ch:large-scale}. For an application-oriented use case, we apply an update period of two months. This period reflects a trade-off between model performance and computational effort, as might be required in practical implementations \cite{Drgona.2020}.
%For further discussion on computational time, we refer to Section \ref{ch:compute}.
We investigate the same methods as in Section \ref{ch:ex_concept_drifts}. Though we neglect pure ALG since eALG performs slightly better and uses less compute. Also, we omit the model from scratch, since the previous studies showed poor performance and high computational effort.

%So far, the experiments considered only 8 buildings that are affected simultaneously by one concept drift.To provide a more robust study, we investigate the most promising methods for 40 buildings and seven years of simulated data with multiple pseudo-random concept drifts generated using Algorithm \ref{algo:data_gen}.

\begin{figure}[t!]
    \centering
    \includegraphics[trim=0.6cm 0 0.6cm 0, clip, width=1 \linewidth]{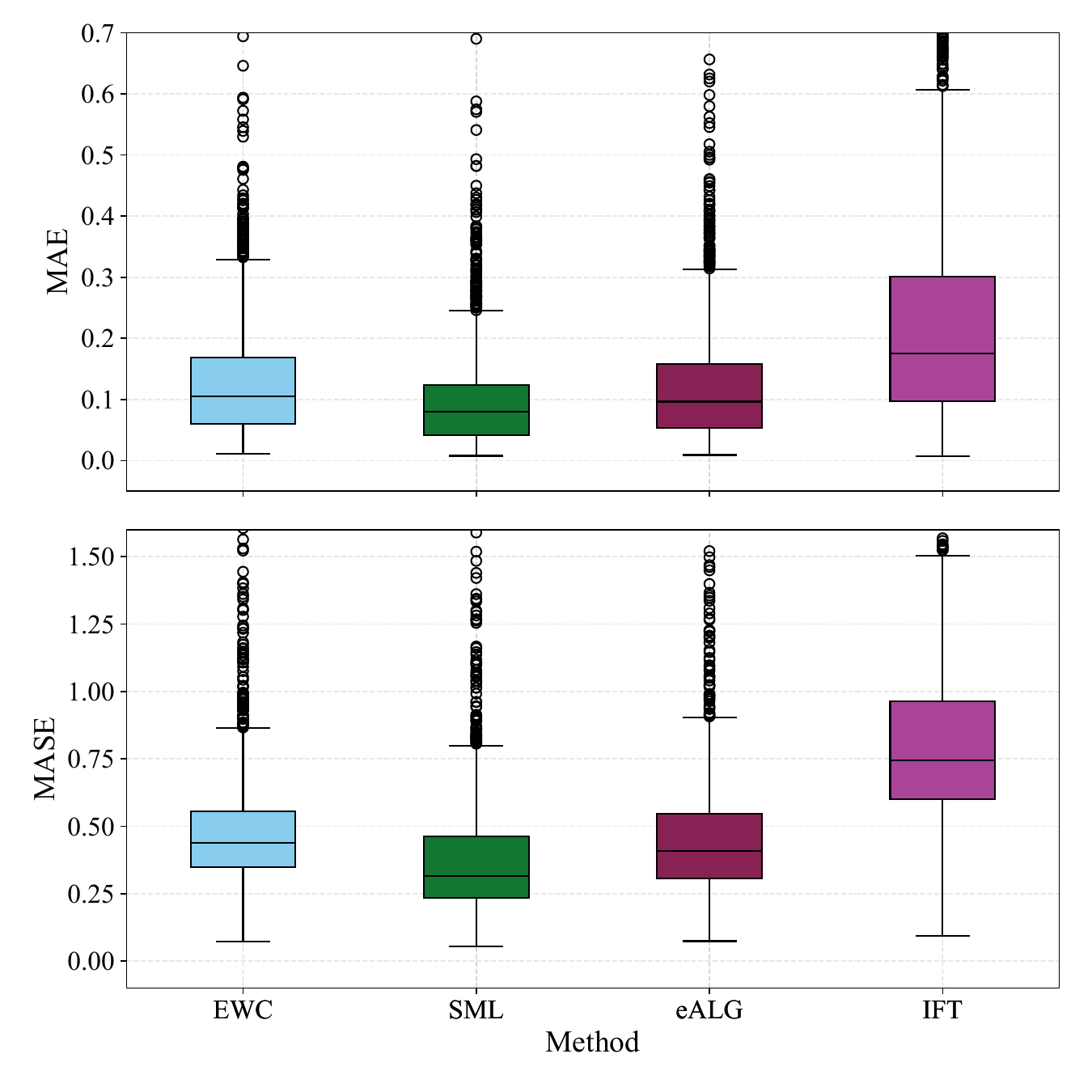}
    \vspace*{-1cm}
    \caption{Box plot of large-scale comparison of different methods shown for \edit{MAE} and MASE values. The data includes the update test error across seven years of 40 buildings with pseudo-random concept drifts generated using Algorithm \ref{alg:data_gen}.}   
    \label{fig:box_plot}
\end{figure}

To present the results, we collect all update test errors for each target and method over the seven-year period. Figure \ref{fig:box_plot} summarizes these results as box plots \edit{for the MAE and MASE values. RMSE and R² values are illustrated in the Appendix Figure \ref{fig:box_plot_2}}. Each box represents the first and third quartiles (25th and 75th percentiles), with the median shown at the 50th percentile. The whiskers extend up to 1.5 times the inter-quartile range, while outliers appear as individual points beyond this range.
%To present the results, we collect each update test error for each individual target and method over seven years. Figure \ref{fig:box_plot} presents the results as box plots. Each box shows the first and third quartiles, corresponding to the 25th and 75th percentiles, respectively. The median is indicated by the 50th percentile. Whiskers extend to 1.5 times the interquartile range beyond the quartiles, and outliers are shown as individual points beyond this range.
Overall, \edit{MAE} and MASE values exhibit similar behavior. SML achieves the lowest mean \edit{MAE (0.098)}, followed by eALG \edit{(0.122)} and EWC \edit{(0.125)}. The benchmark performs worst with a mean of \edit{0.241}, leading to a \edit{48.3}\% (RRI) improvement of SML over IFT.
These results indicate that when multiple drifts occur over time, strategies using smaller amounts of data (SML, EWC) may perform better or similar to those relying on larger datasets (eALG). 
The values for the MASE demonstrate that most of the time, all updating methods beat the naive predictor. Only a few outliers show worse performance. \edit{For IFT, nearly 25\% of the samples perform worse than the naive predictor.}

\newpage
\section{Discussion}
\label{ch:discuss}

This paper investigates adaptive learning during building operation for predicting thermal dynamics. We define adaptive learning as the process of updating models following an initial fine-tuning phase, as new data is acquired over time.
This study focuses on the effects of seasonal variation, building retrofits, and changes in occupancy on the modeling performance. The study employs strategies from Continual Learning (CL), which are Incremental Learning (IL), Elastic Weight Consolidation (EWC), Gradient Episodic Memory (GEM), and a new approach, namely Seasonal Memory Learning (SML). We also compare approaches from Transfer Learning (TL), which are Initial Fine-Tuning (IFT), Incremental Learning on General Model (GIL), Accumulative Learning on General Model (ALG), as well as an adjusted version, which we call event-based Accumulative Learning on General Model (eALG). We consider IFT as the benchmark. Additionally, we employ a model from scratch.
%We conducted experiments over a 5-year building environment with (1) no changes, (2) with a retrofit, and (3) with an occupancy change. Finally, we performed a large-scale analysis in which we compared 40 buildings over seven years, including pseud-random concept drifts.

The first experiment showed the influence of seasonality for adaptive learning, demonstrating that summer data is more difficult to predict than winter data. This is most likely due to a more constant temperature in the winter due to the heat source, compared to more peaks in the summer due to free-floating conditions, as already discussed in Section \ref{ch:ex_undrifted}. An additional observation is that adaptive learning reduces the errors for summer data over time.
%different updating periods on the prediction performance. Results show best performance for an updating period of one month, with little difference to an updating period of two months. The update period of three months showed the worst performance.
%Another finding is that most methods exhibit a downward trend over the years in prediction accuracy. Reasons for this are that more data is available or better adaptation of the models to the target building over time. 
The second experiment analyzed the effect of retrofits and changes in occupancy. The retrofit demonstrated a large effect on the model from scratch and a medium effect on the CL and TL strategies. This is probably due to the generalization capabilities of the general TL model, which profited from pretraining on 450 buildings with different insulation properties. The model trained from scratch could not benefit from prior knowledge and most likely overfitted to the pre-drift state of the target building.
However, the drift reveals that the general TL model appears to perform worse for well-insulated buildings, compared to the broad distribution of buildings. 
\editt{To underline this assumption, we show the average performance of GIL (fine-tune general model for each update step) and IFT (fine-tune general model only once) for undrifted target buildings sorted by U-values in the Appendix Figure \ref{fig:u_values_scatter}. These methods best indicate the general TL models' performance, as these don't follow an recursive updating strategy. GIL and IFT exhibit a U-shaped pattern, achieving their best performance at a U-value of 0.85 $W/(m^2K)$.}
%However, the results in the insulation-level analysis further reinforce the bias of the the general TL model t this observation.
%The reason for this may be the pretraining distribution of the general model, that has a segnificant effect on target distribution performance, as explained in \cite{Felix}.
\editt{The worse performance for well-insulated buildings may result from} \edit{the dataset used for pretraining the general TL model, which employed a uniform distribution of buildings with outer wall U-values ranging from 0.1 to 1.3, as these represent a large distribution of European single-family houses \cite{raisch2025gentlgeneraltransferlearning}. Consequently, well-insulated buildings (U-values around 0.1) represent an edge case within this distribution. Results reveal that the general TL model appears slightly biased toward the mean of the pretrained dataset. Future research could address this limitation by incorporating more buildings that represent the lower end of the U-value range or by developing specialized general models tailored to specific insulation levels, such as those of well-insulated buildings.}
The experiment for the change in occupancy demonstrates little effect on the methods that incorporated the general TL model. This indicates good generalization of the general TL model and the absence of catastrophic forgetting regarding user behavior. The model from scratch, on the other hand, revealed a large effect on the change in occupancy, as it only incorporated data of one occupancy pattern. 
Finally, the large-scale analysis highlights the strong performance of the SML method. SML achieved the best results in \edit{the drifted and undrifted scenarios, followed by eALG.}
%both the large-scale and retrofit experiments. For the occupancy change and the undrifted data, its performance was comparable to eALG, which showed slightly better outcomes.
Overall, the results indicate that plasticity is not observed in thermal dynamics modeling for buildings, as the error remains stable even with extensive training. This may be due to the small number of updates performed over time, compared to traditional CL, which typically involves many more updates. Similarly, catastrophic forgetting is not a major concern, since the focus for buildings is performing well on the new task, unlike traditional CL, which aims to maintain performance on both new and past tasks.

\subsection{Computational resources}
\label{ch:compute}

\begin{figure}[b!]
    \centering
    \includegraphics[width=1 \linewidth]{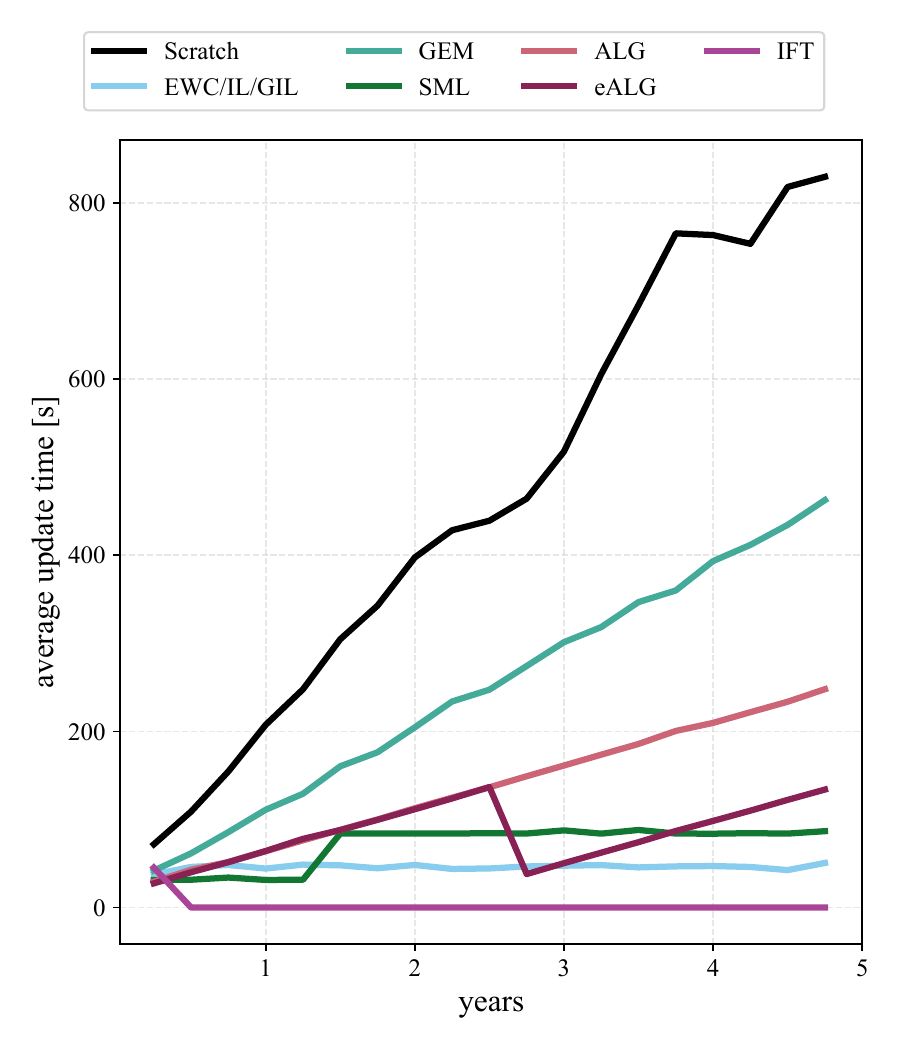}
    \vspace*{-1cm}
    \caption{Average updating time over time for all considered methods. CL methods in green/blue and TL methods in yellow/red.}  
    \label{fig:training_time_plot}
\end{figure}

%We observed differences in computational costs and memory costs for each method. The computational costs are the time required to train the model. The memory costs are the amount of memory required to store the required data for each method. CL focuses in general on using the least amount of memory and computational (training) time while maximizing model performance. TL and a model from scratch do not aim for the tradeoff but rather focus on model performance. 

For real-world deployment, not only model performance but also memory costs and computational time for the updates are crucial. Table \ref{tab:cl_algorithms} shows a comparison regarding the memory costs for all methods using the Big-O notation. A method has memory complexity O(1) if it uses a constant amount of data independent of the update step, and O(N) if its data usage grows proportionally with the update step.

Regarding computational time, we measured the training time for each method and update step, as shown in Figure \ref{fig:training_time_plot}. EWC, GIL, and IL are grouped in the plot to reduce visual clutter, since their trends are similar. 
In general, there is a strong correlation between memory complexity and computational time. O(N) leads to a constant increase in training time. The difference in slope depends on the epochs required for convergence in the training process (Scratch vs ALG and eALG), or the complexity of the method (GEM). GEM constrains gradient updates with information from previous tasks, leading to higher computational demands.
%The model from scratch and GEM exhibit the highest training times, which increase as updates accumulate. The model from scratch processes an increasing amount of training data over time. Also, it requires more epochs to converge due to its random initialization. GEM stores gradient information from previous tasks, leading to higher memory and computational demands. In contrast, ALG and eALG show lower training times. Although they also accumulate data, they require fewer training epochs to converge since they are initialized with the general TL model. 
Event-based ALG neglects old data after an event, leading to less training time afterward. Methods with O(1) maintain constant training time across all update steps, as they store a constant amount of data. Only SML includes a jump after one year, as this is when the method includes previous seasonal data.
%The memory cost follows a similar trend to the training time, as training time primarily depends on the amount of processed data. Table \ref{tab:cl_algorithms} provides a more detailed comparison of data and memory.

\edit{For practical deployment, model updates are typically performed either in the cloud or directly on edge devices \cite{Drgona.2020}. In both cases, minimizing computational time and memory requirements is crucial to maintain low operational costs, making methods with O(1) complexity preferable. Since model updates are only required once every one or two months, the computational burden is generally manageable. Although edge devices are often resource-constrained and training may take longer, convergence can still be achieved as long as sufficient memory is available. A longer update time is acceptable for applications such as control or FDD, where the previous model can still be employed during the retraining process. In contrast, methods with O(N) complexity are better suited for cloud-based retraining, after which updated models can be redeployed to the building system.}

\subsection{Benefit of incoming data}

In the introduction, we posed the research question: ``How does incorporating new measurement data into building dynamics models affect prediction accuracy during operation, and how does the update frequency influence model performance?''
We found that all of the investigated methods lead to a performance increase when incorporating more data. Results revealed a larger increase for the first three years, while stagnating thereafter. 
Accordingly, we would recommend adaptive learning for at least three years, but also for longer if concept drifts occur.
Overall, the investigated methods for adaptive learning outperformed the benchmark of initial fine-tuning by \edit{15-58}\%.
Our experiments demonstrated the best performance for an updating period of one month. Though updating every two months yielded similar results and could therefore be preferred due to computational reasons.
%from CL and TL lead to better performance than IFT. This indicates that after initial fine-tuning should definitely be continued to update the models with incoming data. However, we also found that a model from scratch is not a good alternative. Therefore, we have to contradict the authors of \cite{pinto_sharing_2022}, who suggested using a model from Scratch versus a TL model after one year. They probably came to a different result, as no general TL model was available.

Second, we asked which update strategy is most suitable as more data becomes available, particularly regarding the tradeoff between model accuracy and computational costs. We found that all of the investigated methods from CL and TL lead to better performance than IFT. However, we also found that a model from scratch is not a good alternative. This contradicts the findings of \cite{pinto_sharing_2022}, who suggested using a model from scratch versus a TL model after one year. They probably came to a different result, as no general TL model was available.
% Our experiments revealed the best performance for an updating period of one month. Though updating every two months yielded similar results and could therefore be preferred due to computational reasons.
In general, the methods SML and ALG performed best across the undrifted scenarios. Both methods outperformed the benchmark IFT by around \edit{40}\%. We recommend using SML because it requires significantly less training time and maintains a memory complexity of O(1). If less memory is available, EWC might be a good option. EWC outperformed the benchmark by \edit{28.5}\% for an update period of one month.
%SML performed slightly better when concept drifts occurred. Also, less computational time and memory are required for SML. Therefore, we recommend using our self-invented strategy of SML for building thermal dynamics modeling. If less compute is available, EWC might be a good option.

%“How do drifts in the built environment influence the modeling of building thermal dynamics and which strategies best adapt models to these evolving conditions
Third, we were wondering how drifts influence the modeling and which strategies best adapt models to new conditions.
For seasonal drifts, we observed that summer data is more difficult to predict than winter data. \edit{The best} option to address seasonality is SML. Drifts caused by retrofits and changes in occupancy have a large effect on a model from scratch. The CL and TL strategies were affected less due to the benefits from the general TL model. To be more robust against concept drifts, we recommend using the general TL model in some instance of modeling a target building. In particular, SML performed best among the concept drift study and the large-scale analysis, \edit{closely followed by eALG.}

%How should building dynamics models be updated after initial fine-tuning as more data becomes available over time?
% 1-2 months, CL/TL methods
%How can newly collected data be incorporated into the model in a systematic and effective manner?
%Which update strategy is most suitable as more data becomes available, particularly to capture seasonal shifts in building thermal dynamics?
%How do concept drifts caused by retrofits and occupant changes influence the modeling of building thermal dynamics?
% update period?

\subsection{Limitations and future work}

%One limitation concerns the selection of methods. We acknowledge that many other approaches exist for adaptive learning. However, we tried to focus on a small set of promising methods. The results indicate only small differences between these methods. The type and amount of data appear to have a greater influence. 
There are many approaches to adaptive learning, but necessarily this study could only test a small subset due to computational limitations. However, we focused on a set of promising methods. Future research could employ more advanced methods, especially from the CL community.

Similarly, there are many possible retrofits buildings can undergo. This study tested only a retrofit of the whole building envelope because this leads to a large change in dynamics. Retrofits of individual components (e.g., windows) are also possible. Future research could address other types of drifts, such as changes in heating equipment, as mentioned in Section \ref{ch:concept_drifts}.

Another limitation concerns the updating strategy. This study uses a periodic, time-based update scheme, which is commonly applied. Future work could investigate alternative strategies, such as updates triggered by performance degradation.

\edit{The model architecture also presents limitations. While we employed a recurrent architecture suited for time series forecasting, more advanced models, such as Transformers, could potentially enhance prediction accuracy. However, this could lead to bottlenecks during retraining for on-site hardware. Exploring such alternatives in an adaptive learning setting could be part of future research.}
%However, our focus is not on achieving the best possible performance, but rather on gathering insights on how to employ data over time. Future research can focus on more advanced model architecture and can explore the use of other pretrained models such as TimesNet

Finally, we used simulated data in this study. Simulations allow controlled concept drifts and multi-year data generation across different buildings. \edit{Future work should focus on collecting multi-year real-world data that is affected by concept drifts and includes all necessary feature variables for thermal dynamics modeling to evaluate the proposed methods. This data will serve as an assessment for prediction accuracy in an offline manner. Subsequently, the thermal dynamics model should be tested in pilot applications for control or fault detection and diagnosis. Such studies will strengthen the methodology’s robustness and address the sim-to-real transfer challenge.}
\section{Conclusion}
\label{ch:conclusion}

In this paper, we presented a comprehensive comparison of Continual Learning (CL), Transfer Learning (TL), and training from scratch for adaptive modeling of building thermal dynamics. The methods were evaluated in an undrifted setting to assess seasonal performance and under concept drifts caused by retrofits and occupancy changes.

The results show that CL and TL strategies outperform initial fine-tuning and models trained from scratch. 
We recommend using the general pretrained TL model from \cite{raisch2025gentlgeneraltransferlearning} as a starting point for modeling, as it demonstrates resilience to concept drifts during adaptive learning.
Furthermore, we introduced Seasonal Memory Learning (SML), a method that incorporates both newly available data and data from the same season one year ago. Across experiments, SML achieved the highest prediction accuracy on both undrifted and drifted data while maintaining low computational cost. We therefore recommend SML as an effective strategy for adaptive modeling of building thermal dynamics. With these findings, we hope to advance energy efficiency in long-term building operation.

\section*{CRediT authorship contribution statement}

\textbf{Fabian Raisch}: Conceptualization, Software, Methodology, Investigation, Writing - Original Draft, Writing - Review \& Editing
\textbf{Max Langtry}: Conceptualization, Methodology, Writing - Original Draft, Writing - Review \& Editing
\textbf{Felix Koch}: Software, Writing - Review \& Editing
\textbf{Ruchi Choudhary}: Supervision, Writing - Review \& Editing
\textbf{Christoph Goebel}: Supervision, Conceptualization, Writing - Review \& Editing
\textbf{Benjamin Tischler}: Supervision, Conceptualization

\section*{Declaration of competing interests}

The authors declare that they have no known competing financial interests or personal relationships that could have appeared to influence the work reported in this paper.

\section*{Data availability}

The code and data used to perform the experiments in this study are available at \cite{CLGit}.

\section*{Acknowledgements}

Max Langtry is supported by the Engineering and Physical Sciences Research Council, through the CDT in Future Infrastructure and Built Environment: Resilience in a Changing World, Grant [EP/S02302X/1].

\appendix
\section{Additional results}
\label{app:results}

\begin{table}[ht]
\caption{Averaged yearly \edit{MAE} values across all target buildings for different updating methods with a model updating period of one month and no concept drifts in the building.}
\label{tab:yearly_rmse}
\centering
\renewcommand{\arraystretch}{1.2}
\vspace{0.2cm}
\begin{tabular}{lccccc}
\hline
Method & Year 1 & Year 2 & Year 3 & Year 4 & Year 5 \\
\hline
Scratch & 0.5193 & 0.1333 & 0.1194 & 0.1113 & 0.1059 \\
IL & 0.0869 & 0.0826 & 0.0784 & 0.0752 & 0.0799 \\
EWC & 0.0853 & 0.0798 & 0.0754 & 0.0735 & 0.0766 \\
GEM & 0.0843 & 0.0823 & 0.0787 & 0.0767 & 0.0816 \\
SML & 0.0862 & \textbf{0.0630} & \textbf{0.0536} & \textbf{0.0512} & \textbf{0.0533} \\
GIL & \textbf{0.0824} & 0.0833 & 0.0863 & 0.0879 & 0.0925 \\
ALG & 0.0893 & 0.0694 & 0.0630 & 0.0633 & 0.0636 \\
IFT & 0.1218 & 0.1267 & 0.1305 & 0.1306 & 0.1365 \\
\hline
\end{tabular}
\end{table}

\begin{table}[t]
\caption{ \editt{Prediction errors (MAE, RMSE, MASE, R²) averaged over five years for different updating methods with noisy data. Noise was added according to \cite{choi_performance_2023}.}}
\label{tab:noise}
\centering
\vspace{0.2cm}
\begin{tabular}{{lcccc}}
\hline
\textbf{Method} & MAE & RMSE & MASE & R² \\
\hline
Scratch & 0.380 & 0.500 & 0.850 & -1.997 \\
IL & 0.287 & 0.366 & 0.658 & 0.120 \\
EWC & 0.286 & 0.365 & 0.656 & 0.144 \\
GEM & 0.286 & 0.365 & 0.657 & 0.130 \\
SML & \textbf{0.280} & \textbf{0.352} & \textbf{0.644} & \textbf{0.291} \\
GIL & 0.291 & 0.381 & 0.667 & 0.061 \\
ALG & 0.281 & 0.354 & 0.645 & 0.235 \\
IFT & 0.314 & 0.406 & 0.714 & 0.217 \\
\hline
\end{tabular}
\end{table}

% alternatively use \rotatebox{90}{<content>}
\begin{comment}
\begin{table*}[!b]
    \centering
    \begin{tabular}{c@{\hskip .33cm}|ccccccc| cc} 
        \rot{Target building} & \rot{$U_{wall}$ $[W/(m^2K)]$} & \rot{\textbf{$c_{wall}$ $[kJ/(m^2K)]$}} & \rot{\textbf{$f_{win}$}}  & \rot{\textbf{$A_{ground}$ $[m^2]$}} & \rot{$T_{sp,\,day}$ [\textdegree C] } & \rot{$\Delta T_{night}$ [\textdegree C]} & \rot{Weather} & \rot{MAE} & \rot{RMSE} \\  \hline
        T1 & 0.25 & 40  & 0.16 & 70 & 22.0 & 1.0 & Bratislava & 0.159972 & 0.318152\\
        T2 & 0.25 & 280 & 0.19 & 100 & 21.0 & 0.0 & Amsterdam & 0.136382 & 0.293101  \\
        T3 & 0.55 & 150 & 0.16 & 70& 23.0 & 0.0 & Amsterdam & 0.072861 & 0.177894\\
        T4 & 0.55 & 280 & 0.19 & 100 & 20.5 & 1.5 & Munich & 0.086462 & 0.227841 \\ 
        T5 & 0.85 & 40 & 0.16 & 70 & 22.0 & 2.5 & Munich & 0.049821 & 0.090908\\ 
        T6 & 0.85 & 150 & 0.19 & 100 & 22.5 & 0.5 & Bratislava & 0.066239 & 0.133945 \\ 
        T7 & 1.15 & 280 & 0.16 & 70 & 23.0 & 0.0  & Bratislava & 0.066305 & 0.162694\\ 
        T8 & 1.15 & 40 & 0.19 & 100 & 23.0 & 1.5 & Amsterdam  & 0.059759 & 0.161862 \\ \hline
     \end{tabular}
     \caption{Properties of target buildings, selected to cover distribution from Table \ref{tab:targets_data}, according to \cite{raisch2025gentlgeneraltransferlearning}. $T_{sp,\,day}$ is the daytime temperature setpoint and $\Delta T_{night}$ a potential night setback.}
     \label{tab:8targets_errors}
\end{table*}
\end{comment}

\newcommand{\fit}{\footnotesize\it}
\begin{table*}[t]
\caption{\edit{Comparison of averaged prediction errors (RMSE, MASE, MAE, R²) with 95\% confidence interval over five years across update periods.}}
\label{tab:grouped_avg_errors}
\centering
\vspace{0.2cm}
\begin{tabular}{{l|cc|cc|cc}}
\hline
\multirow{2}{*}{\textbf{Method}} & \multicolumn{2}{|c|}{1 month-update} & \multicolumn{2}{|c|}{2 month-update} & \multicolumn{2}{|c}{3 month-update} \\
& \fit RMSE & \fit MASE & \fit RMSE & \fit MASE  & \fit RMSE & \fit MASE \\
\hline
Scratch & 0.330 $\pm$ 0.040 & 1.029 $\pm$ 0.105 & 0.367 $\pm$ 0.073 & 1.090 $\pm$ 0.175 & 0.397 $\pm$ 0.104 & 1.160 $\pm$ 0.247 \\
IL & 0.168 $\pm$ 0.007 & 0.483 $\pm$ 0.020 & 0.171 $\pm$ 0.010 & 0.515 $\pm$ 0.031 & 0.179 $\pm$ 0.012 & 0.555 $\pm$ 0.045 \\
EWC & 0.165 $\pm$ 0.007 & 0.474 $\pm$ 0.020 & 0.163 $\pm$ 0.009 & 0.467 $\pm$ 0.025 & 0.170 $\pm$ 0.012 & 0.488 $\pm$ 0.034 \\
GEM & 0.172 $\pm$ 0.008 & 0.483 $\pm$ 0.019 & 0.177 $\pm$ 0.010 & 0.519 $\pm$ 0.030 & 0.182 $\pm$ 0.012 & 0.549 $\pm$ 0.046 \\
SML & \textbf{0.133} $\pm$ 0.007 & \textbf{0.359 }$\pm$ 0.017 & \textbf{0.136} $\pm$ 0.009 & \textbf{0.373} $\pm$ 0.028 & \textbf{0.142} $\pm$ 0.012 & \textbf{0.392} $\pm$ 0.037 \\
GIL & 0.196 $\pm$ 0.009 & 0.514 $\pm$ 0.019 & 0.203 $\pm$ 0.012 & 0.554 $\pm$ 0.030 & 0.215 $\pm$ 0.015 & 0.629 $\pm$ 0.053 \\
ALG & 0.145 $\pm$ 0.007 & 0.423 $\pm$ 0.019 & 0.143 $\pm$ 0.009 & 0.403 $\pm$ 0.023 & 0.146 $\pm$ 0.010 & 0.406 $\pm$ 0.030 \\
IFT & 0.231 $\pm$ 0.010 & 0.747 $\pm$ 0.030 & 0.298 $\pm$ 0.032 & 0.884 $\pm$ 0.070 & 0.339 $\pm$ 0.046 & 1.009 $\pm$ 0.114 \\
\hline
& \fit MAE & \fit R² & \fit MAE & \fit R²  & \fit MAE & \fit R² \\
\hline
Scratch & 0.197 $\pm$ 0.031 & -0.715 $\pm$ 0.931 & 0.219 $\pm$ 0.058 & -0.541 $\pm$ 1.008 & 0.236 $\pm$ 0.079 & -0.530 $\pm$ 0.982 \\
IL & 0.081 $\pm$ 0.004 & 0.867 $\pm$ 0.022 & 0.088 $\pm$ 0.007 & 0.913 $\pm$ 0.020 & 0.095 $\pm$ 0.009 & 0.927 $\pm$ 0.022 \\
EWC & 0.078 $\pm$ 0.004 & 0.877 $\pm$ 0.021 & 0.080 $\pm$ 0.006 & 0.920 $\pm$ 0.018 & 0.085 $\pm$ 0.008 & 0.932 $\pm$ 0.021 \\
GEM & 0.081 $\pm$ 0.004 & 0.838 $\pm$ 0.028 & 0.088 $\pm$ 0.006 & 0.885 $\pm$ 0.030 & 0.092 $\pm$ 0.008 & 0.904 $\pm$ 0.033 \\
SML & \textbf{0.061} $\pm$ 0.004 & \textbf{0.928} $\pm$ 0.015 &\textbf{0.064} $\pm$ 0.006 & \textbf{0.947} $\pm$ 0.016 & \textbf{0.070} $\pm$ 0.009 & \textbf{0.960} $\pm$ 0.013 \\
GIL & 0.087 $\pm$ 0.005 & 0.720 $\pm$ 0.053 & 0.094 $\pm$ 0.006 & 0.752 $\pm$ 0.074 & 0.106 $\pm$ 0.010 & 0.833 $\pm$ 0.062 \\
ALG & 0.070 $\pm$ 0.004 & 0.910 $\pm$ 0.016 & 0.069 $\pm$ 0.005 & 0.940 $\pm$ 0.015 & 0.071 $\pm$ 0.007 & 0.953 $\pm$ 0.016 \\
IFT & 0.129 $\pm$ 0.008 & 0.798 $\pm$ 0.031 & 0.199 $\pm$ 0.031 & 0.881 $\pm$ 0.020 & 0.231 $\pm$ 0.044 & 0.905 $\pm$ 0.017 \\
\hline
\end{tabular}
\end{table*}

\begin{figure*}[]
    \centering
    \includegraphics[width=1 \linewidth]{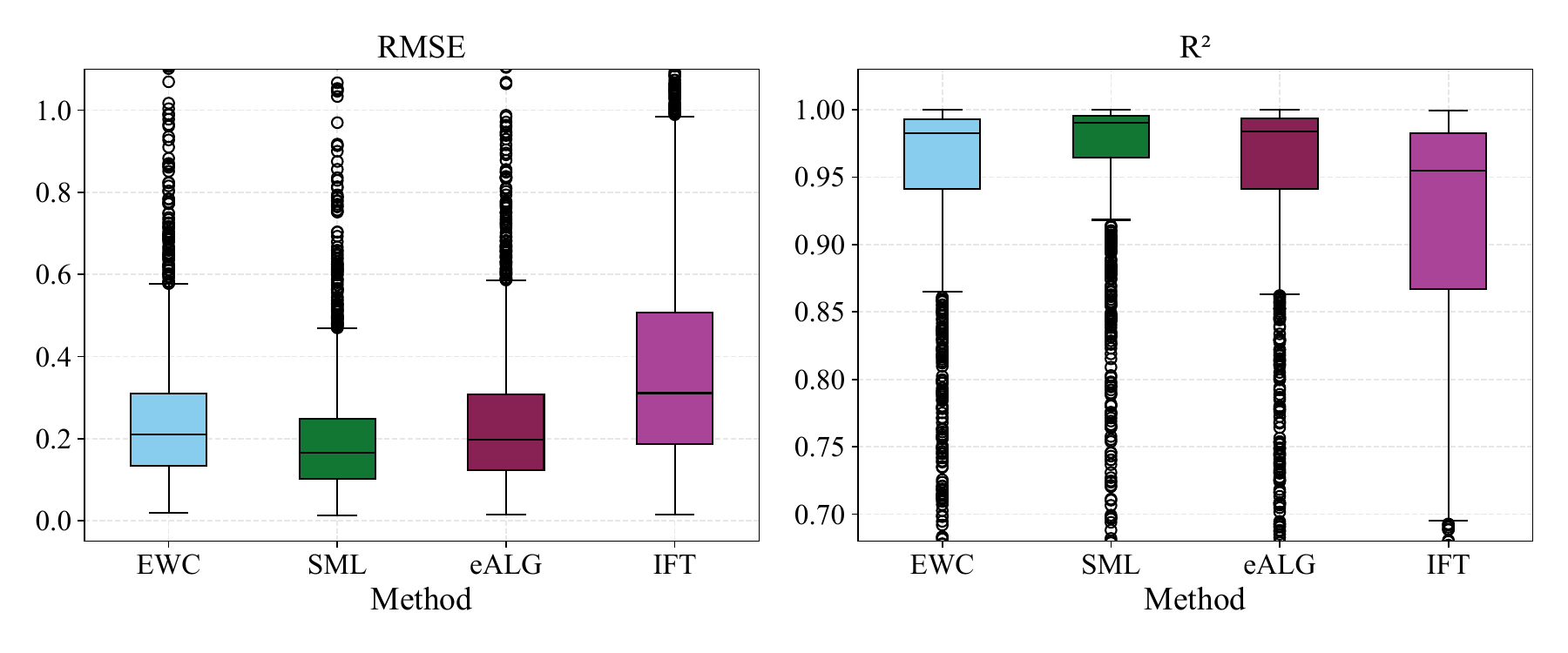}
    \vspace*{-1cm}
    \caption{\edit{Box plot of large-scale comparison of different methods shown for RMSE and R² values. The data includes the update test error across seven years of 40 buildings with pseudo-random concept drifts generated using Algorithm \ref{alg:data_gen}.}}   
    \label{fig:box_plot_2}
\end{figure*}

\begin{figure}[]
    \centering
    \includegraphics[trim=0.7cm 0 0.5cm 0, clip, width=\linewidth]{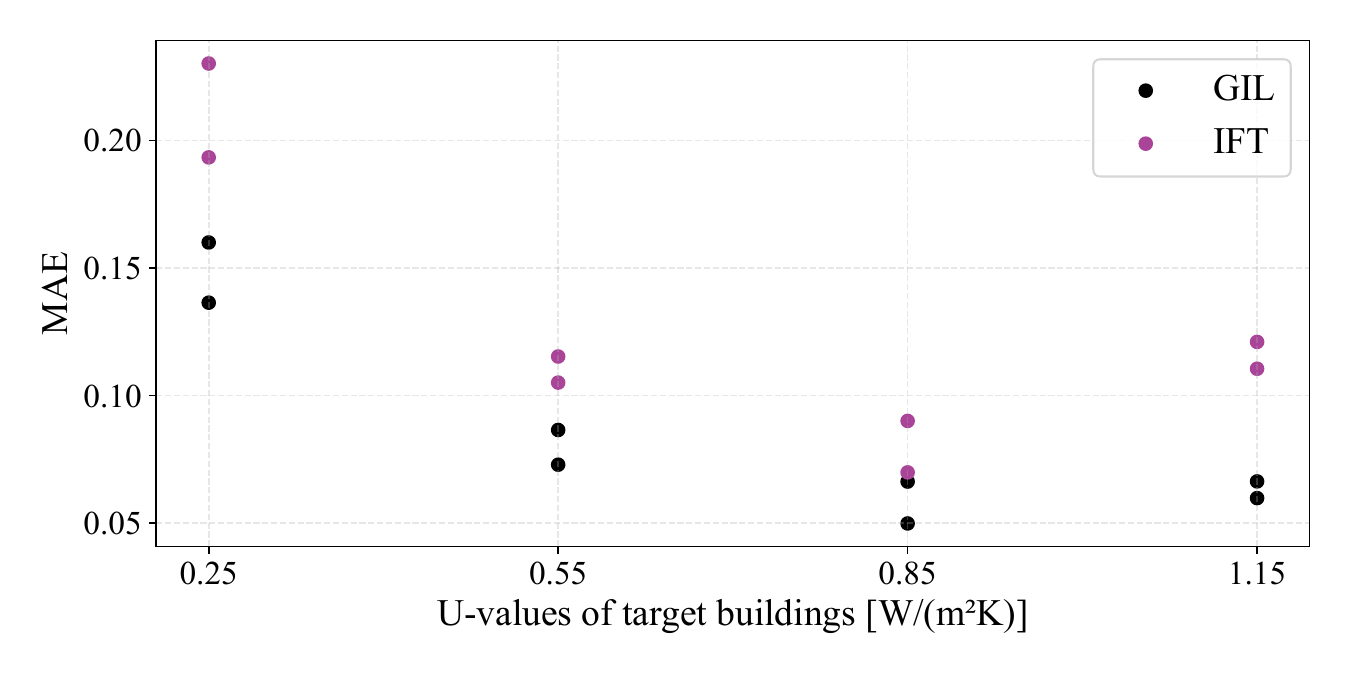}
    \vspace*{-0.7cm}
    \caption{\editt{MAE values for updating methods GIL and IFT for each target building from Table \ref{tab:8targets} sorted according to U-values (better insulation corresponds to a smaller U-value). Note: only two buildings fall into each U-value category, limiting the reliability of these observations. For more comprehensive studies, we refer to \cite{builDaReview, pinto_sharing_2022, builda2}.}}
    \label{fig:u_values_scatter}
\end{figure}

%\clearpage
\bibliographystyle{elsarticle-num} 
\bibliography{ref.bib}



%\begin{thebibliography}{00}

%% For numbered reference style

\end{document}